\documentclass[prx,twocolumn,showpacs,superscriptaddress,floatfix, nofootinbib]{revtex4-2}
\usepackage{amsmath,amssymb,amsfonts,mathrsfs, amsbsy}
\usepackage{graphicx}
\usepackage{tikz}
\usepackage[bookmarks=true, colorlinks=true, linkcolor=blue, urlcolor=blue, citecolor=blue, bookmarks=true, hyperindex=true]{hyperref}

\usepackage[T1]{fontenc}
\usepackage{upgreek}
\usepackage{xcolor}
\usepackage[normalem]{ulem}
\usepackage{multirow}
\usepackage{float}
\usepackage{array}
\usepackage{caption}
\usepackage{comment}
\newcommand{\dd}{\mathrm{d}}
\usepackage{diagbox}

\makeatletter
\long\def\@makecaption#1#2{%
  \par\vskip\abovecaptionskip
  \begingroup
  \leftskip\z@
  \rightskip\z@
  \parfillskip\z@
  \noindent\textbf{#1.} #2\par
  \endgroup
  \vskip\belowcaptionskip}
\makeatother

\usepackage{array}  
\usepackage{lipsum} 

\newcommand{\be}{\begin{equation}}
\newcommand{\ee}{\end{equation}}

\newcommand{\beq}{\begin{eqnarray}}
\newcommand{\eeq}{\end{eqnarray}}

\def\H1{\widehat{H}_1}

\newcommand{\ket}[1]{\left| #1 \right>}
\newcommand{\bra}[1]{\left< #1 \right|}

\usepackage{xcolor} 

\begin{document}

\title{
Multiscale Structure of Eigenstate Thermalization\\
}



\author{Pavel Orlov, Rustem Sharipov, and Enej Ilievski\\[1ex]
\small Faculty for Mathematics and Physics, University of Ljubljana, Jadranska 19, SI-1000 Ljubljana, Slovenia}

\begin{abstract}

The eigenstate thermalization hypothesis provides a framework for 
understanding thermalization in isolated quantum many-body systems by characterizing statistical properties of local observables in energy eigenstates.
Here we demonstrate that distributions of matrix elements in macroscopic systems may depend not only on the macrostate parameters, such as the densities of local conserved charges, but generally also on the properties of ensembles used in sampling eigenstates. To this end, we depart from the conventional analysis of microcanonical windows and consider statistical ensembles with an adjustable scale parameter prescribing the magnitude of charge fluctuations. We specifically consider an integrable field theory that permits efficient numerical sampling of matrix elements and reliable extrapolation to the thermodynamic limit. Moreover, in this system, we identify a class of states that enables explicit closed-form computation of the suppression rate of matrix elements.
Our findings reveal an underlying multiscale structure of matrix elements captured by a non-analytic fluctuation-scale dependence of algebraic exponents governing their statistical properties.

\end{abstract}

\maketitle

\section{Introduction}

The notion of thermal equilibrium is the bedrock principle of statistical mechanics which encapsulates the idea that large many-body systems, in spite of their microscopic complexity, evolve toward macroscopic states that are characterized by a small set of thermodynamic variables. Understanding how such universal behavior emerges from time-reversible laws of the underlying microscopic dynamics is a fundamental problem in statistical physics. The \emph{eigenstate thermalization 
hypothesis} (ETH)~\cite{Deutsch91, Srednicki94} provides a versatile theoretical framework to explain various aspects of thermalization phenomena. The core idea of ETH is that, upon restricting to only \emph{local} measurements, generic quantum systems already exhibit thermal properties at the level of individual energy eigenstates, thereby providing a microscopic justification for the validity of equilibrium statistical ensembles.

In the past decades, the ETH has been extensively studied in the domain of quantum lattice models with local interactions \cite{Rigol_2008, Rigol_2009, Ikeda_2011, Steinigeweg_2013, Sorg_2014, Khodja_2015, Fratus_2015, Mondaini_2016, Beugeling_2014, Steinigeweg_2014, Kim_2014, LeBlond2020,Dymarsky_2018} (see also a review \cite{D_Alessio_2016}), in quantum field theories~\cite{Lashkari:2016vgj, He_2017, Basu_2017, Datta_2019,Delacretaz_2023, Srdinek_2024, Fukushima_2023} and lately also in quantum circuits~\cite{Chan_2019,Hahn_2024, Fritzsch_2021, Fritzsch:2024ppm}. In recent years, the theoretical toolbox for exploring thermalization and related phenomena has expanded beyond the scope of standard ETH, including the ETH approach to thermalization in classical systems~\cite{orlov2026} and the so-called `full ETH' and its relation to the theory of free probability \cite{Mingo2017} which deals with the structure of higher-order correlation functions~\cite{Foini_2019, Pappalardi_2022,PappalardiFelix,PappalardiFelixPRb,ClayesFull-1,Clayes-2,camargo2025free,Junhe2025}. Meanwhile, various mechanisms that lead to violation of ETH have been investigated as well \cite{Moudgalya_2022,Avdoshkin,FadingLev, Sierant_2025,Lev_Fabian,Lev_scaling}.

While standard numerical diagonalization methods, or more recent approaches based on tensor networks \cite{Banuls-diagonal,Banuls-diagonal2,Luo_2024}, provide indispensable tools for studying ETH in quantum lattice systems, finite-size effects and large statistical fluctuations may sometimes preclude reliable extrapolation to the thermodynamic limit. A controlled analytic approach is therefore essential for obtaining a unified and comprehensive understanding, which motivates to look for simple, and possibly exactly solvable, models. In this respect, integrable systems are most natural candidates, as they enable a rich variety of analytical approaches.
Indeed, it is well established that local observables in integrable quantum systems do undergo thermalization in the sense that the long-time averages are accurately described by generalized Gibbs ensembles~\cite{Vidmar_GGE,Enej-complete-GGE,Enej-quasi-charges}. This statement relies on the fact that diagonal matrix elements obey the so-called diagonal ETH adapted to the presence of higher conservation laws~\cite{Enej-complete-GGE,Enej-quasi-charges, Dymarsky_2019, Ishii_2019}. However, due to the intricate structure of eigenstates, the off-diagonal matrix elements in integrable systems turn out to be much more nuanced to describe \cite{LeBlond_2019,essler-LiebLiniger,rottoli2025eigenstate}. 

\begin{figure}[h!]
    \centering
    \vspace{-10pt}
     \includegraphics[width=1.0\columnwidth]{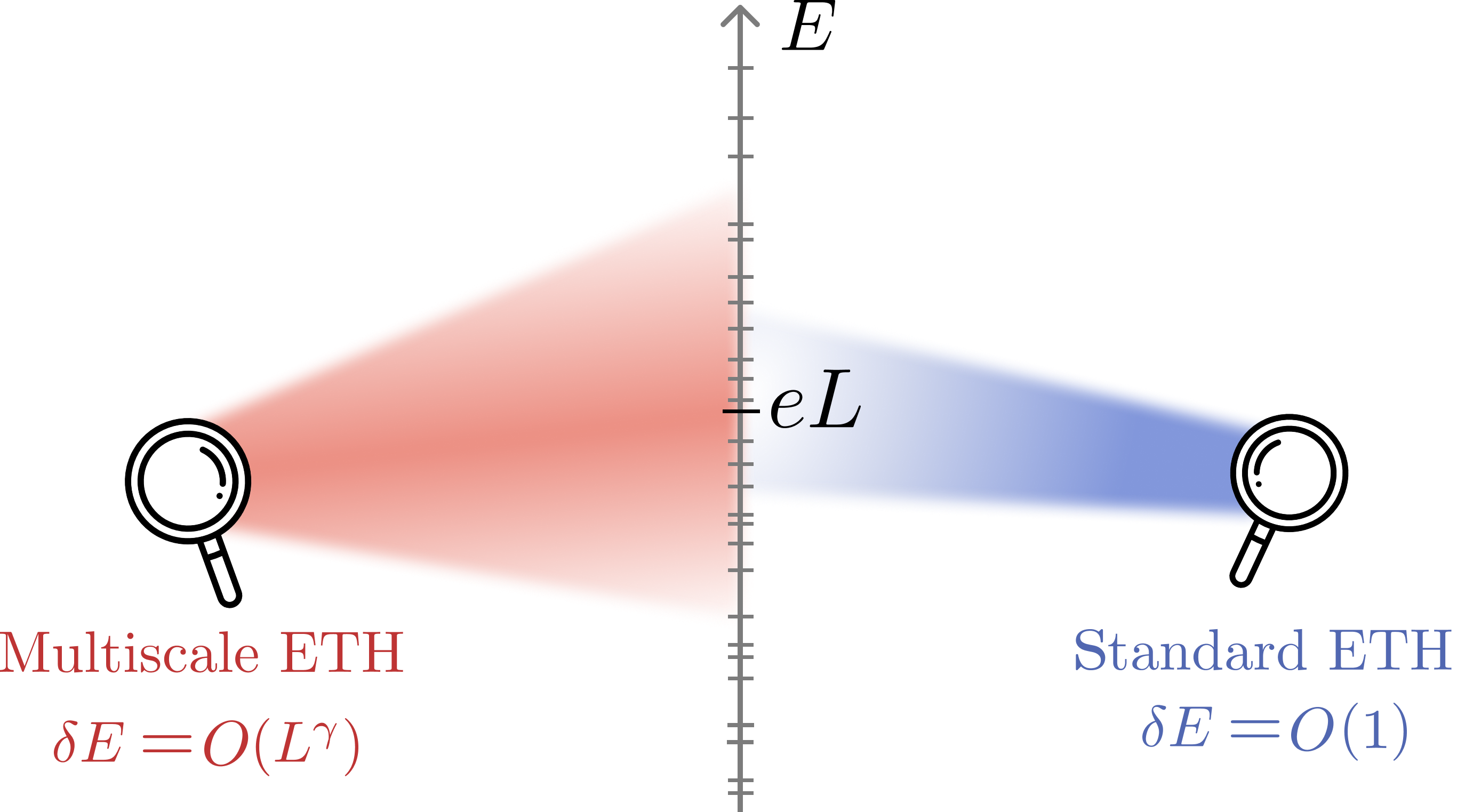}
     \caption{{\bf Probing multiscale structure of eigenstate thermalization.} The standard ETH scenario concerns the structure of local observables in the Hamiltonian eigenbasis by averging over narrow energy windows of width $\delta E = O(1)$. With aid of the \emph{multiscale} analysis one can access the structure across parametrically larger energy scales set by the magnitude of fluctuations $\delta E = O(L^{\gamma})$, with an adjustable fluctuation exponent $\gamma$.}
     \label{fig:multiscale}
\end{figure}

One specific aspect of ETH that we aim to address in this work, and that has so far received virtually no attention, is the dependence of the matrix element distributions on the choice of the averaging ensemble. Statistical properties of matrix elements are conventionally probed in microcanonical ensembles of eigenstates confined to finite energy windows of width $\delta E = O(L^{0})$. One may, however, think of ensembles with larger energy fluctuations, e.g. $\delta E = O(L^{\gamma})$ with a characteristic fluctuation scale $\gamma \in [0,1]$, as illustrated in Fig.~\ref{fig:multiscale}. For instance, $\gamma = 1/2$ is the scale of typical energy fluctuations in canonical (Gibbs) ensembles, whereas the largest available scale $\gamma = 1$ involves eigenstates at different energy densities, which thus belong to different macrostates. Such fluctuation scales emerge rather naturally in quantum quench protocols, where the system is typically initialized in a state with large, yet subextensive, energy fluctuations $\delta E = O(L^{\gamma})$ with $\gamma \in [0,1)$. Therefore, understanding the behavior of matrix elements at larger energy separations may play an essential role in the ensuing dynamics. While in generic chaotic systems the nearby matrix elements at scale $\gamma = 0$ are expected to follow the standard ETH predictions, in the extreme case of distinct macrostates at $\gamma = 1$ one may ask whether suppression of matrix elements becomes parametrically larger, signaling a breakdown of the standard ETH ansatz. This intuition readily motivates the following question: \emph{do the statistical properties of matrix elements depend on the fluctuation scale} $\gamma$?

\begin{figure}[h!]
    \centering
     \includegraphics[width=0.80\columnwidth]{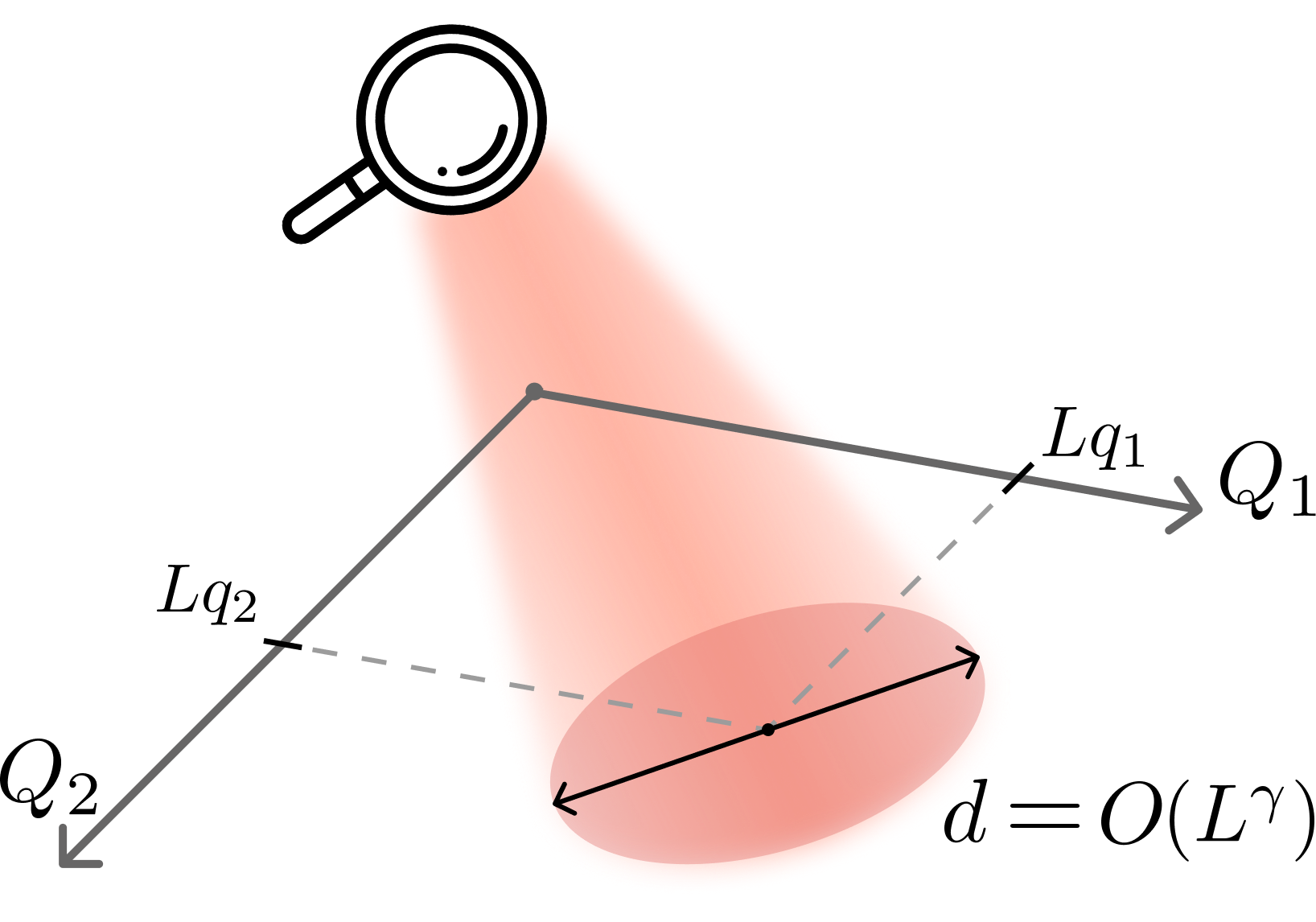}
     \caption{{\bf Introducing the distance.} Schematic visualization of a higher-dimensional space of charge eigenvalues $Q_{i}$ (for brevity presented here for two charges, $N_{Q}=2$). Red shaded region pertains to an ensemble of eigenstates with the typical \emph{distance} (charge separation) of the order $d=O(L^{\gamma})$ residing within a macrostate characterized by charge densities $q_{i}$.}
     \label{fig:intro2}
\end{figure}

In this work, we aim to address the above aspect in a systematic fashion
by characterizing distributions of off-diagonal matrix elements at all available fluctuation scales. For this purpose, our goal is to extract the algebraic scaling exponents $p(\gamma)$ and $q(\gamma)$ that quantify the asymptotic properties of the distribution mean and width, respectively, depending on $\gamma$.
Moreover, by introducing an appropriate distance between eigenstates, our approach is not only applicable to chaotic many-body systems but also accommodates systems with an arbitrary number of local conservation laws, including integrable systems, see Fig.~\ref{fig:intro2}.

For a successful realization of the proposed program, it is instrumental to not only gain access to large system sizes but also to develop an efficient numerical sampling procedure. These requirements make it challenging to implement our program in the domain of chaotic systems, mainly due to the limitations of diagonalization techniques. To facilitate the analysis, we instead consider here an integrable quantum field theory of a massless bosonic field which permits a remarkably efficient computation of matrix elements for a certain class of local observables. Remarkably, in this model, we derived an explicit analytical result for the suppression rate of matrix elements and the corresponding scaling exponent $p(\gamma)$ for the entire range of fluctuation scales. Moreover, within our framework, we construct ensembles with a tunable magnitude of charge fluctuations and devise a protocol for sampling eigenstates from them, enabling numerical extraction of the scaling exponents $p(\gamma)$ and $q(\gamma)$ up to system sizes that are comparable to spin chains of lengths $L\sim 10^6$.

Our analysis uncovers a non-trivial and non-analytic $\gamma$-dependence of the matrix elements statistics, exposing an underlying multiscale structure of matrix elements in equilibrium macrostates. At the same time, our findings reveal how the presence of the higher conserved charges plays a crucial role in shaping the statistical properties of off-diagonal matrix elements in integrable systems, while clarifying the origin of the anomalous structure previously observed in other integrable models~\cite{essler-LiebLiniger, rottoli2025eigenstate}.
 
\paragraph*{Outline.}
The paper is structured as follows. To set the stage, we begin in Sec.~\ref{sec:ETH_intro} by a short presentation of the basic notions of the standard ETH in ergodic quantum lattice systems, followed by a brief discussion of integrable models in Sec.~\ref{sec:ETH_integrable}.
We proceed in Sec.~\ref{sec:beyondETH} by introducing the key ingredients of our program for studying the multiscale structure of ETH.
The main results of our study, alongside a short exposition of the model, are succinctly summarized in Sec.~\ref{sec:summary}, which we wrap up in Sec.~\ref{sec:discussion} with a brief discussion and comparison with the previous relevant works. The remainder of the paper is devoted to various technical aspects related to the model (its spectrum and its macrostates) and also include detailed information on the employed computational algorithms. In Sec.~\ref{sec:conclusion} we reiterate our main findings and mention several important directions for future work.

\tableofcontents

\section{Eigenstate Thermalization Hypothesis}
\label{sec:ETH_intro}

\subsection{ETH in chaotic quantum systems}

The eigenstate thermalization hypothesis (ETH) asserts that, in the thermodynamic limit, matrix elements of local observables in the energy eigenbasis acquire a particular universal form resembling that of random Gaussian matrices~\cite{Deutsch91,Srednicki94,D_Alessio_2016}.

In one-dimensional Hamiltonian lattice systems without additional conserved quantities, \emph{diagonal} ETH states that the diagonal matrix elements $A_{mm}$ of a \emph{local} observable $\hat{A}$, written in the eigenbasis of a many-body Hamiltonian $\hat{H}$, $\hat{H}\ket{m} = E_m \ket{m}$, do not depend on microscopic details but only on macroscopic properties of thermodynamic eigenstates,
\begin{equation}\label{eqn:diagonal_ETH}
    A_{mm} = \mathcal{A}(e) + o(1),
\end{equation}
where $e$ denotes the energy density, $E_m = e\,L + o(1)$, and $L$ is the system size. An important corollary of Eq.~\eqref{eqn:diagonal_ETH} is that the long-time average of $\hat{A}$ matches the corresponding microcanonical value for physically relevant initial states.

A more general statement --- known as the \emph{off-diagonal} ETH~\cite{Srednicki_1999} --- stipulates that typical off-diagonal matrix elements between two eigenstates at the same energy density $e$ are exponentially small in $L$, with a suppression rate set by the thermodynamic entropy density $s(e)$. This statement proves useful for bounding temporal fluctuations of observables during equilibration dynamics.

Aside from the overall scale, off-diagonal matrix elements are expected to behave as pseudo-random variables whose statistical properties depend smoothly on the energy separation. The off-diagonal ETH ansatz is traditionally presented in the form
\begin{equation}\label{standard-ETH}
    A_{mn} = e^{-s(e)L/2} f_{A}(e, \omega) R_{mn}.
\end{equation} 
Here, the matrix element is taken between states $\ket{m}$ and $\ket{n}$ at energy density $e$ and finite energy separation $\omega_{mn} \in [\omega, \omega +\delta\omega]$ with $\omega = O(1)$ and small $\delta\omega$. The so-called `spectral function' $f_{A}(e,\omega)$ is assumed to be a smooth function of $\omega$, while the numbers $R_{mn}$ represent the residual pseudo-randomness. Statistical properties of $R_{mn}$ are obtained after the uniform average $\mathbb{E}_{\omega, \delta \omega}$ over all pairs of states with an energy difference $\omega_{mn} \in [\omega, \omega + \delta \omega]$. Their first two moments are given by
\begin{equation}
    \mathbb{E}_{\omega,\delta\omega}[R_{mn}] = 0, \quad \mathbb{E}_{\omega, \delta \omega}[ |R_{mn}|^2 ] = 1,
\end{equation}
while higher moments are ultimately related to the free probability theory~\cite{Foini_2019, Pappalardi_2022} and are tied to the non-commutative nature of the algebra of observables. 

The spectral function $f_{A}(e ,\omega)$ is indeed none other than a dynamical two-point correlation function $C_{A}(t) \equiv \langle \hat{A}(t) \hat{A}(0)\rangle_{\beta} -  \langle \hat{A}^2 \rangle_{\beta}$ in the frequency (Fourier) domain, computed in a thermal Gibbs state $\hat{\rho}_{\beta} = Z_{\beta}^{-1} e^{-\beta \hat{H}}$, 
\begin{equation}
    C_{A}(t) = \int_{-\infty}^{+\infty} d \omega \,e^{i \omega t} e^{\beta \omega /2} |f_{A}(e_{\beta} , \omega)|^{2},
\end{equation}
with $e_{\beta}$ denoting the average energy density corresponding to $\hat{\rho}_{\beta}$. The higher-point correlators are similarly encoded in the higher moments of matrix elements~\cite{Foini_2019,Pappalardi_2022}.

\subsection{ETH in integrable systems}
\label{sec:ETH_integrable}

The standard form of ETH can be directly extended to systems with multiple local conservation laws $\{\hat{Q}^{(k)}\}_{k =1}^{N_Q}$. In principle, it can also be adapted to \emph{integrable} systems that host an extensive number $N_{Q}\sim L$ of (quasi)\emph{local} charges. In doing so, one has to be careful with the proper identification of macrostates; an unambiguous determination of a macrostate requires (besides energy density $q^{(1)} \equiv e $) additional information about the complete set of independent (quasi)local conserved quantities $\hat{Q}^{(k)}$. Macrostates are therefore uniquely specified by providing the complete set of charge densities $\{ q^{(k)} \}_{k}$. The mechanism that underlies conventional thermalization towards local maximum-entropy states is indeed unaffected by the presence of additional conservation laws provided one selects an appropriate macrostate (leading to the notion of the generalized Gibbs ensembles \cite{Vidmar_GGE,Enej-quasi-charges}). The suitably generalized version of the diagonal ETH accordingly reads
\begin{equation}\label{diagonal-ETH-integrable}
    A_{mm} = \mathcal{A}(\{q^{(k)}\}) + o(1).
\end{equation}
As already corroborated in earlier studies of integrable systems~\cite{Cassidy_2011, Dymarsky_2019, Essler_2016, essler-LiebLiniger, Ishii_2019}, the long-time averages of local observables generically equilibrate to ensemble averages associated with such macrostates.

Thus, at the conceptual level at least, there is no fundamental distinction between chaotic and integrable systems as far as the diagonal ETH is concerned: diagonal matrix elements in the thermodynamic limit retain only the coarse-grained information encoded in the macrostate, while microscopic details of individual eigenstates are washed out in the thermodynamic limit.
The main challenge in practice lies in identifying the complete set of parameters specifying the macrostate.

In contrast, validity of the \emph{off-diagonal} ETH ansatz in integrable systems is considerably more nuanced and remains the topic of active research~\cite{LeBlond_2019, LeBlond2020, LeBlond2020-2, LeBlond-2020-3, Zhang_2022, essler-LiebLiniger, rottoli2025eigenstate}. 
By leveraging the Bethe-ansatz techniques, the recent studies of the Lieb–Liniger model~\cite{essler-LiebLiniger} and the Heisenberg spin chain~\cite{rottoli2025eigenstate} numerically investigated matrix elements of certain local observables for system sizes far beyond the reach of conventional approaches. In these studies, the authors report a scaling form of the type $A_{mn} \sim \exp\!\left(- c_A L \log L - L M_{mn}\right)$, indicating an anomalous, parametrically enhanced, suppression 
compared to the standard ETH scenario which predicts exponential suppression, $A_{mn}= e^{-O(L)}$. Moreover, the effective randomness, encoded in pseudo-random variables $M_{mn}$, appears at the subleading order $e^{-O(L)}$, contrasting the $O(1)$ randomness of the ETH ansatz. Finally, $M_{mn}$ were shown to follow a Gumbel distribution instead of the log-normal distribution commonly associated with the ETH.

This brings us to a subtle yet crucial point: the sampling procedure used in Refs.~\cite{essler-LiebLiniger, rottoli2025eigenstate} only restricts the energy separation between eigenstates (or finitely many charges), letting the higher charges fluctuate freely. For this reason, a direct comparison with the ETH predictions is not meaningful. In contrast, the generalized diagonal ETH~\eqref{diagonal-ETH-integrable} indicates that an appropriate distance between eigenstates should take into account the complete set of (quasi)local conserved quantities. As we detail out in Sec.~\ref{sec:beyondETH}, integrable systems permit us to introduce a suitable distance (see Eqs.~\eqref{distance-int} and \eqref{distance-Young}) which we utilize in this work to investigate statistical properties of matrix elements in a certain integrable model.
In addition, in Sec.~\ref{sec:summary} we explain the origin of the observed anomalous scaling reported in~\cite{essler-LiebLiniger, rottoli2025eigenstate}.

\paragraph*{Sparse versus dense observables.}
The off-diagonal ETH is conventionally concerned with local observables supported on a compact subsystems which connect an entropic, i.e. exponential in $L$, number of eigenstates. Such typical observables, which can be referred to as \emph{dense}, are to be contrasted with so-called \emph{sparse} observables which only yield a polynomially growing number of non-zero matrix elements. The latter are commonly seen in non-interacting theories, where one typically studies observables that comprise a finite number of quasiparticles creation and annihilation operators. To facilitate a fair comparison between chaotic and integrable systems it is therefore mandatory to pick a dense observable.

\section{Multiscale structure of matrix elements}
\label{sec:beyondETH}

In this section we outline a general program for studying statistical properties of off-diagonal matrix elements that can be implemented on a wide variety of quantum models, including both chaotic and integrable systems.

\subsection{Probing different fluctuation scales}

\paragraph*{Distance between eigenstates.}
In Hamiltonian chaotic systems, assuming absence of additional conservation laws, the energy difference $\omega_{nm}=E_{n}-E_{m}$ provides a natural distance between two arbitrary eigenstates $\ket{n}$ and $\ket{m}$,
\begin{equation}
    \text{dist}(n,m) = |\omega_{mn}|.
\end{equation}
Accordingly, the ensemble-averaged distance, denoted by $\omega$, is the only ``thermodynamic coordinate'' of the standard ETH ansatz, Eq.~\eqref{standard-ETH}, for the target energy density $e$ that specifies the macrostate.

In the case of multiple commuting local charges $\{\hat{Q}^{(k)} \}_{k =1}^{N_Q}$, each charge $\hat{Q}^{(k)}$ can in principle be assigned a separate coordinate $\omega^{(k)}$. Evidently, such a multivariate extension of the ETH posits a rather formidable computational task, especially for integrable systems. A more feasible approach is to perform a simpler univariate analysis, which accounts for all local charges on equal footing. This can be easily achieved by introducing a single `global' distance parameter, for example, the 1-norm distance on the macrostate manifold, 
\begin{equation}\label{distance-int}
    \text{dist}(n,m) = \frac{1}{N_{Q}}\sum_{k=1}^{N_Q} |Q_{n}^{(k)} - Q_{m}^{(k)}|.
\end{equation}
Although the same distance is meaningful even in integrable systems with $N_{Q}=O(L)$, there might exist more suitable options. Particularly, since eigenstates in integrable models are characterized by a unique set of quantum numbers (most often linked with quantized momenta of quasiparticle excitations), it is more convenient to define a distance between eigenstates in terms of those quantum numbers (see Sec.~\ref{section-macrostates} and Eq.\eqref{distance-Young} for a concrete example).

\paragraph*{Fluctuation scale.}
Next, we consider a one-parameter family of statistical ensembles $\mathcal{E}_{\gamma}$ characterized by the following properties:
\begin{equation}
\begin{aligned}
   Q^{(k)} &\equiv  \langle \hat{Q}^{(k)} \rangle_{\mathcal{E}_{\gamma}} = L\,q^{(k)} + o(L), \\
   \delta Q^{(k)} &\equiv \sqrt{\langle ( \hat{Q}^{(k)} )^2 \rangle_{\mathcal{E_{\gamma}}} - \langle \hat{Q}^{(k)}\rangle^2_{\mathcal{E}_{\gamma}}} = O(L^{\gamma}),
\end{aligned}
\end{equation}
In other words, while charge averages scale extensively, the corresponding typical fluctuations are required to grow with $L$ algebraically with exponent $\gamma \in[0,1]$, which we refer to as the \emph{fluctuation scale}. Note that for any $\gamma<1$, eigenstates from $\mathcal{E}_{\gamma}$ have equal charge densities $q^{(k)}$ at large $L$, and therefore all such ensembles correspond to the same macrostate (fully characterized by $\{ q^{(k)} \}$).
The extreme case of $\gamma=1$, on the other hand, corresponds to ensembles with extensively large charge fluctuations involving eigenstates from different macrostates.
Importantly, the fluctuation scale can be alternatively controlled via the mean distance between eigenstates sampled from $\mathcal{E}_{\gamma}$, 
\begin{equation}
    \langle{\rm dist}(n,m)\rangle_{\mathcal{E}_{\gamma}} = O(L^{\gamma}).
\end{equation}

We also emphasize that exponent $\gamma$ alone does not uniquely specify a statistical ensemble. One option for chaotic systems (with only energy conservation) is to pick a two-parameter family of `microcanonical' ensembles characterized by the energy density $e$ and window of width $d=O(L^{\gamma})$:
\begin{equation}\label{micro-ensemble}
    \hat{ \rho }_{\rm mc} (e, d ) \simeq \sum_{n: \text{ } |E_n- eL|<d } \ket{n} \bra{n}.
\end{equation}
In integrable systems, one can analogously define the `generalized' microcanonical ensemble by restricting all the charges to lie in some (formally infinite-dimensional) hypershell~\cite{Essler_2016, Ishii_2019}. 

To our knowledge, all the previous studies of the off-diagonal ETH in integrable systems account only for the energy separation (or finitely many charges) among eigenstates, mostly considering the microcanonical energy ensemble, Eq.~\eqref{micro-ensemble} with $d=O(1)$, or the canonical Gibbs ensemble. Due to fluctuations of higher charges, the fluctuation scale in such ensembles is actually given by $\gamma=1/2$ (and not by $\gamma = 0$ as in chaotic energy-conserving systems). As discussed in Sec.~\ref{sec:discussion}, this fact has important consequences for the properties of off-diagonal matrix elements.

 \subsection{Statistics of matrix elements}

Our central objective is to quantify statistical properties of off-diagonal matrix elements on different fluctuation scales. Given a local observable $\hat{A}$, we are particularly interested in the probability distributions of the pseudo-random variables\footnote{This representation conveniently avoids dealing with exponentially small numbers. Moreover, in the case of exponential suppression of $|A_{mn}|^{2}$, $\kappa_{mn}$ represent intensive quantities.}
\begin{equation}
    \kappa_{mn} = -\frac{1}{L} \text{log}|A_{mn}|^2.
\end{equation}
We mainly focus on the mean value and variance of $\kappa_{mn}$ with eigenstates $\ket{m}$ and $\ket{n}$ sampled from ensemble $\mathcal{E}_{\gamma}$,
\begin{equation}\label{suppresion-width}
\begin{aligned}
    &\kappa \equiv \langle \kappa_{mn} \rangle_{\mathcal{E}_{\gamma}}, \\
    &\delta \kappa \equiv \sqrt{ \langle \kappa_{mn}^2 \rangle_{\mathcal{E}_{\gamma}} - \langle \kappa_{mn} \rangle^2_{\mathcal{E}_{\gamma}} },
\end{aligned}
\end{equation}
and their large-$L$ behavior on different fluctuation scales set by $\gamma$.
The mean value $\kappa$ physically represents the average \emph{suppression rate} of $|A_{mn}|^{2}$, while $\delta \kappa$ corresponds to the \emph{distribution width} measuring the typical magnitude of randomness.

In order to determine how $\kappa$ and $\delta \kappa$ scale asymptotically with system size $L$, we introduce two \emph{algebraic scaling exponents}
\begin{equation}\label{scaling-exponents}
    p(\gamma) \equiv \frac{\partial \text{log}(\kappa)}{\partial \text{log}(L)}, \qquad q (\gamma) \equiv \frac{ \partial \text{log}(\delta \kappa)}{ \partial \text{log}(L) }.
\end{equation}
Note that the ETH ansatz, Eq.~\eqref{standard-ETH}, assumes $p_{\rm ETH} = 0$ and $q_{\rm ETH} = -1$.

Additionally, we will also study the full probability distribution function (PDF) of $\kappa_{mn}$. Our objective is to examine how the latter depends on the fluctuation scale or on the other properties of ensembles $\mathcal{E}_{\gamma}$.

In order to reliably extract the algebraic scaling exponents $p(\gamma)$ and $q(\gamma)$, it is pivotal to pick a system that permits an efficient computation of eigenstates and sampling from ensembles $\mathcal{E}_{\gamma}$ for large system sizes. For this reason, we carry out our program in an integrable model.

\section{Main results and Discussion}
\label{sec:summary}

Here, we succinctly summarize the main results of our work and compare our findings to the previous studies.

\subsection{Scaling of matrix elements in an integrable model}

Although integrability offers a rich variety of analytical tools, including e.g. the Bethe ansatz algebraic construction of eigenstates and explicit representations of matrix elements~\cite{Korepin1982, Slavnov1989,Korepin1993,Kitanine_1999}, there nevertheless remain several stumbling blocks: (i) solving the Bethe ansatz equations poses a nontrivial computational problem on its own which severely restricts the accessible system sizes; (ii) there is no obvious algorithm to perform efficient sampling of eigenstates from ensembles with different fluctuation scales.

There indeed exists a simple integrable system that meets the required criteria -- a
\emph{quantum Benjamin--Ono hierarchy} residing inside the theory of a massless bosonic field ~\cite{Alba_2011,Nazarov_2013}. The key feature of this model is that the full spectrum of eigenstates can be explicitly constructed in terms of integer partitions (Young diagrams) see Eqs.~\eqref{Schur-eigenstates} and \eqref{eigenvalues}.
Furthermore, matrix elements of certain operators, the so-called vertex operators, are explicitly known for any pair of eigenstates~\cite{Alba_2011}, see Eq.\eqref{matrix-element}. 
More importantly, we were able to identify a simple class of states we call the ``staircase diagrams'' (see Fig.~\ref{staircase-diagrams}), for which we were able to obtain an analytical result for the suppression rate $\kappa$ as a function of the eigenstate distance. 

This model, as well, enables us to efficiently generate eigenstates from ensembles with desired fluctuation scales $\gamma$. In our implementation, we employ the canonical Gibbs state, also known in the literature as the \textit{Vershik ensemble}, which has been extensively studied before~\cite{vershik-stat,FreimanVershikYakubovich1997,Yakubovich2001,Funaki_2013}. Crucially, fluctuations in this ensemble have been explicitly described and are governed by a Gaussian process of magnitude $O(L^{1/2})$.
We then modify such Vershik ensembles by rescaling the magnitude of fluctuations with an adjustable parameter $d$ while keeping its structure intact. In order to probe the fluctuation scale $\gamma$, we impose $d=O(L^{\gamma})$. In such a setting, $d$ is proportional to the distance \eqref{distance-Young} on the space of Young diagrams. As we detail in Sec.~\ref{sec:Mod_Vershik}, such a procedure enables an efficient sampling of matrix elements.

\subsubsection{Analytical results for the staircase diagrams}

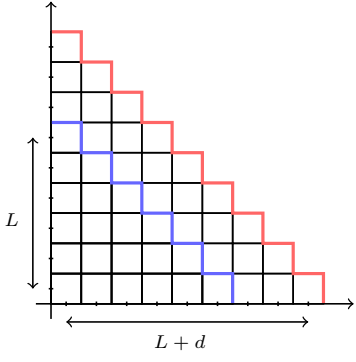
\begin{figure}[H]
\centering
\begin{tikzpicture}[scale=0.4]

\def\partitionSmall{6,5,4,3,2,1}
\def\partitionLarge{9,8,7,6,5,4,3,2,1}

\foreach \length [count=\row from 0] in \partitionLarge {
    \pgfmathtruncatemacro{\n}{\length}
    \foreach \col in {0,...,\numexpr\n-1} {
        \draw[semithick] (\col,\row) rectangle ++(1,1);
    }
}

\foreach \length [count=\row from 0] in \partitionSmall {
    \pgfmathtruncatemacro{\n}{\length}
    \foreach \col in {0,...,\numexpr\n-1} {
        \draw[semithick] (\col,\row) rectangle ++(1,1);
    }
}

\draw[very thick, red!60]
(9,0)
-- (9,1)
-- (8,1)
-- (8,2)
-- (7,2)
-- (7,3)
-- (6,3)
-- (6,4)
-- (5,4)
-- (5,5)
-- (4,5)
-- (4,6)
-- (3,6)
-- (3,7)
-- (2,7)
-- (2,8)
-- (1,8)
-- (1,9)
-- (0,9);

\draw[very thick, blue!60]
(6,0)
-- (6,1)
-- (5,1)
-- (5,2)
-- (4,2)
-- (4,3)
-- (3,3)
-- (3,4)
-- (2,4)
-- (2,5)
-- (1,5)
-- (1,6)
-- (0,6);

\draw[->, semithick] (-0.5, 0) -- (10, 0);   
\draw[->, semithick] (0, -0.5) -- (0, 10);   


\def\maxi{8.5}
\def\maxj{8.5}

\foreach \x in {0,...,\maxi} {
    \draw[semithick] (\x+0.5, 0.075) -- ++(0,-0.15);
}

\foreach \y in {0,...,\maxj} {
    \draw[semithick] (0.075,\y+0.5) -- ++(-0.15,0);
}

\node at (-1.3, 2.8) { \scriptsize $L$ };
\draw[<->,semithick] (-0.6,0.5) -- (-0.6, 5.5);

\node at (4.25, -1.3) { \scriptsize $L+d$ };
\draw[<->,semithick] (0.5,-0.6) -- (8.5, -0.6);

\end{tikzpicture}

\caption{{\bf Staircase diagrams.} The family of Young diagrams of the type
$\boldsymbol{\lambda}_{L} = (L, L-1, ... , 1)$, called \emph{staircase diagrams}, is ideally suited for an analytical study of the matrix-elements suppression rate. A diagram at the distance $d$ away from $\boldsymbol{\lambda}_{L}$ marked by the blue limit shape is simply the diagram $\boldsymbol{\lambda}_{L+d} = (L+d, L+d-1, ..., 1)$ marked by the red limit shape.}

\label{staircase-diagrams}

\end{figure}

The staircase diagrams, depicted in Fig.~\eqref{staircase-diagrams}, enable us to derive an explicit analytic formula for the asymptotic behavior of the matrix elements as a function of distance.
In the thermodynamic limit, the suppression rate $\kappa$ becomes a function of a single scaling variable,
\begin{equation}\label{scaling_variable}
    u = \frac{d^2}{L} \log{\left(\frac{L}{d}\right)},
\end{equation}
where $d$ here is proportional to the distance between two staircase diagrams; specifically, it exhibits a simple, linear form
\begin{equation}\label{analytic_result}
    \kappa = \kappa_0 + \frac{1}{2} u, 
\end{equation}
modulo the terms subleading in $L$, where $\kappa_0$ is an observable-dependent constant (see Sec.~\ref{sec:staircasediagrams} for a more detailed discussion).

\begin{figure}[t]
    \centering
     \includegraphics[width=1.0\columnwidth]{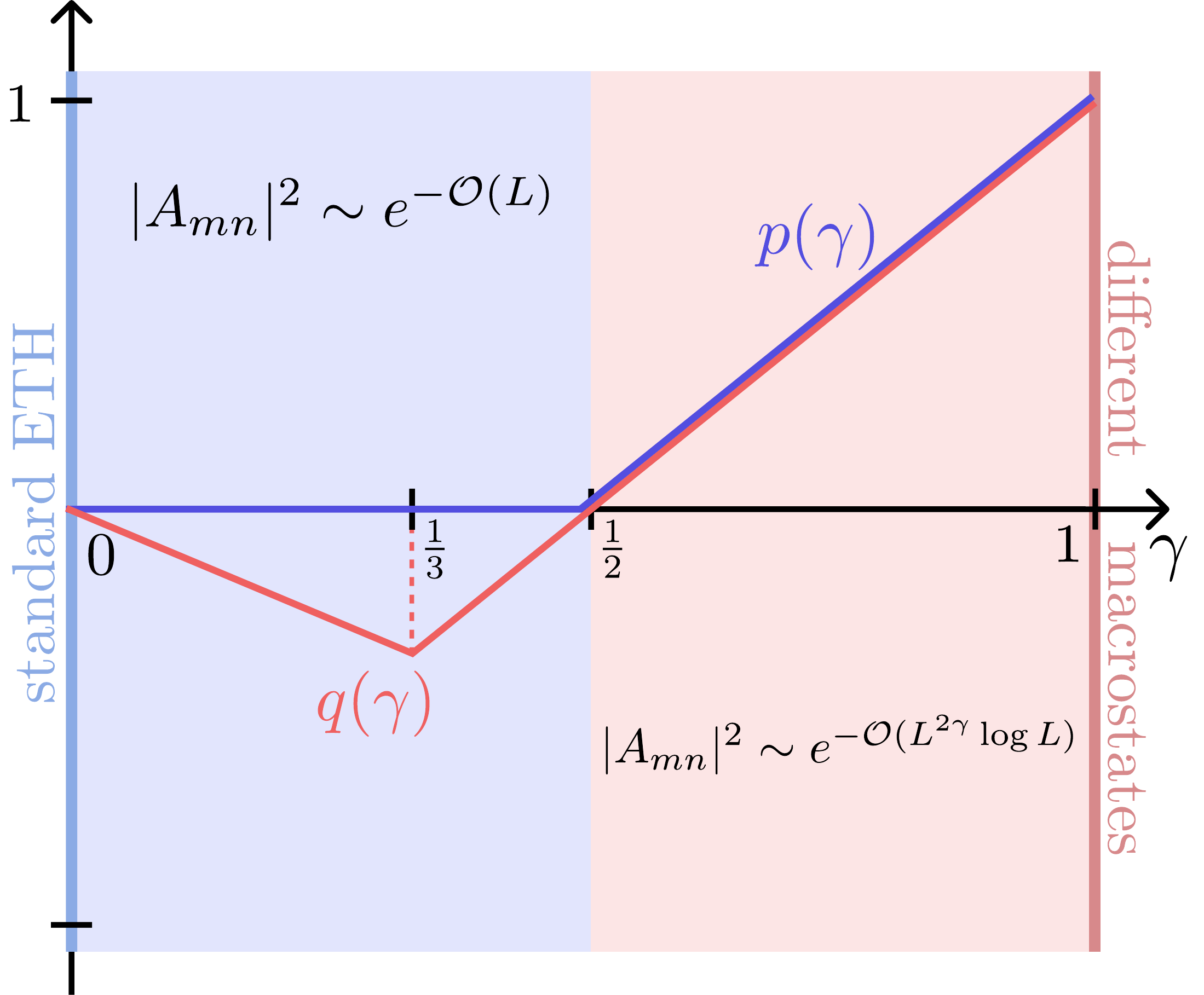}
     \caption{{\bf Fluctuation scale dependence of matrix elements statistics in the modified Vershik ensembles.} The algebraic scaling exponents $p(\gamma)$ (blue) and $q(\gamma)$ (red) quantifying the asymptotic growth of the distribution mean $\kappa$ and width $\delta \kappa$, respectively, as functions of the fluctuation scale $\gamma$. The scaling exponent $p(\gamma)$ governing the suppression rate of $|A_{mn}|^{2}$ becomes parametrically enhanced beyond the thermal characteristic scale $\gamma=1/2$, with an additional multiplicative logarithmic correction.}
     \label{fig:scaling-exponents-plot}
\end{figure}

The scaling form \eqref{analytic_result} greatly simplifies the extraction of the algebraic scaling exponent $p(\gamma)$, Eq.~\eqref{scaling-exponents}. By imposing $d = O(L^{\gamma})$, it immediately follows that variable $u$ vanishes for $\gamma<1/2$, whereas for $\gamma\geq 1/2$ it diverges for large $L$ as $u=O(L^{2\gamma-1}\log{L})$. Therefore
\begin{equation}\label{kappa_scalings}
\begin{aligned}
    \kappa &= O(1) \quad &{\rm for}&\quad 0\leq \gamma <1/2,\\
    \kappa &= O(L^{2\gamma-1}\log L) \quad &{\rm for}&\quad 1/2\leq \gamma < 1. \\
\end{aligned}
\end{equation}
Additionally, we find that for $\gamma=1$, i.e. the case of different macrostates, $\kappa = O(L)$ (but without a logarithmic correction). Together, this implies $p(  \gamma \geq 1/ 2 )=2\gamma-1$ and $p(\gamma<1/2)=0$, see Fig.~\ref{fig:scaling-exponents-plot}.

\subsubsection{Scaling exponents in modified Vershik ensembles}

By numerically investigating the scaling properties of $\kappa(L,d)$ in modified Vershik ensembles, we find that $u$ is still an adequate scaling parameter, cf. Eq.~\eqref{scaling_variable}. In particular, $\kappa$ satisfies the scaling law 
\begin{equation}\label{eq:scaling_mod_k}
\boxed{
    \kappa=f_{\kappa} \left(\frac{d^2}{L}\log\left(\frac{L}{d}\right) \right).}
\end{equation}
In distinction from the staircase diagrams in Eq.~\eqref{analytic_result}, the scaling function $f_{ \kappa}(u)$ exhibits linear asymptotics only at large $u$, $f_{\kappa} \sim u$, while for small $u$ it saturates to a constant value $\kappa_0$.
By repeating the previous analysis, we thus arrive at the same scaling laws of the suppression rate, Eq.~\eqref{kappa_scalings}.

Contrary to $\kappa$, we lack a theoretical prediction for the scaling behavior of the distribution width $\delta \kappa$. We nonetheless empirically find that the distribution width follows the scaling form
\begin{equation}
    \boxed{\delta \kappa = L^{-1/3} f_{\delta \kappa}(d/L^{1/3}),}    
\end{equation}
with the scaling function $f_{\delta \kappa}(v)$ obeying the asymptotics $f(v)\sim v^{2}$ and $f(v) \sim v^{-1}$ at large and small $v$ respectively. The above scaling form implies $q(\gamma \leq 1/3) = -\gamma$ and $q(\gamma > 1/3) = 2\gamma - 1$.

In summary, numerically extracted scaling exponents in the modified Vershik ensembles read as follows:
\begin{equation}\label{scaling-exp-numerics}
\begin{aligned}
    p(\gamma) = 
    \begin{cases}
        0, \quad &\gamma <1/2, \\
        2 \gamma -1 , \quad &\gamma\geq1/2, \\
    \end{cases} \\
    q(\gamma) = \begin{cases}
        - \gamma , \quad &\gamma <1/3, \\
        2\gamma-1, \quad &\gamma \geq 1/3.
    \end{cases}
\end{aligned}
\end{equation}
A detailed analysis of scaling exponents is performed in Sec.~\ref{sec:scaling-exp}.

\subsubsection{Distributions of matrix elements in modified Vershik ensembles}

Apart from the scaling exponents, we also analyzed the full probability distribution of matrix elements. At the ``thermal'' fluctuation scale $\gamma = 1/2$, we find excellent agreement with the Gumbel 
distribution, compatibly with recent results obtained in other integrable systems~\cite{essler-LiebLiniger, rottoli2025eigenstate}. 

By extending this analysis to other fluctuation scales, we find that distributions remain universal and of the Gumbel form for all $\gamma > 1/3$. On the other hand, in the regime $\gamma<1/3$ we observe systematic deviations from the Gumbel form, and we instead find the statistics to be well approximated by the skew-normal distribution, see Sec.~\ref{sec:distributions}.

\subsection{Discussion}
\label{sec:discussion}

\subsubsection{Comparison with ETH studies in chaotic models}

Although no previous work has investigated the regime $\gamma > 0$, we can nevertheless attempt to infer the scaling exponent $p$ directly from the large-frequency asymptotics of the ETH spectral function
by expressing the ETH suppression rate, Eq.~\eqref{standard-ETH}, in the form
\begin{equation}\label{kappa-ETH}
\kappa_{\text{ETH} } = s(e) - \frac{1}{L} \log |f_A(e,\omega)|^2.
\end{equation}

While there is strong evidence \cite{D_Alessio_2016,LeBlond_2019,Rabinovici_2022,Cirac-ETH} that $|f_{A}(e,\omega)|$ falls off exponentially in generic chaotic systems, $|f_A(\omega)| \sim e^{-O(\omega)}$,
there might exist systems which violated that, e.g. Refs.~\cite{Lev-Holstein,Lev_Marcin} found a stronger suppression of the form $|f_A(\omega)| \sim e^{-O(\omega^2)}$. In the former case, Eq.~\eqref{kappa-ETH} predicts $|A_{mn}| \sim e^{-O(L)}$ for all fluctuation scales $0\leq \gamma \leq 1$. In the latter case, one however retrieves $p(\gamma)=2\gamma-1$, precisely as in our model (albeit without the additional logarithmic correction in the suppression rate $\kappa$).

While the same comparison could in principle be made for exponent $q$ that characterizes the distribution width, we are not aware of any previous study of the frequency dependence of distribution widths $\delta \kappa$ in the domain of chaotic systems.

It should be emphasized that in our model, even though the scaling exponent $p(0)=0$ plays well with the ETH prediction, the suppression rate $\kappa_0$ does not equal the entropy density of the corresponding macrostate. Moreover, $\kappa_0$ is found to depend explicitly on an observable. This feature is presumably related to the underlying integrability.

\subsubsection{Comparison with ETH studies in integrable models}

As already emphasized earlier, all the previous studies of ETH in integrable systems effectively probe the ``thermal'' fluctuation scale $\gamma=1/2$ owing to fluctuations of higher charges. At this particular value, our results are well-aligned with the recent studies~\cite{essler-LiebLiniger, rottoli2025eigenstate} which investigated the structure of matrix elements of certain local observables in the Lieb–Liniger Bose gas and in the Heisenberg spin-$1/2$ chain by sampling thermal eigenstates. In these works, the authors numerically extracted the following scaling form for the matrix elements,
$|A_{mn}|^{2} \sim \exp(-c_A L \log L - M_{mn}^A L)$,
with the pseudo-random numbers $M_{mn}^A$ being approximately Fréchet or Gumbel distributed. Apart from confirming the conjectured form, we here explain the appearance of a somewhat mysterious logarithmic correction using a fully analytical treatment of the staircase diagrams. Moreover, Refs.~\cite{essler-LiebLiniger, rottoli2025eigenstate} also analyzed the suppression rate for different macrostates, finding a stronger suppression of the form $e^{-O(L^2)}$, which again aligns with our findings.

In summary, the `anomalous' behavior observed in Refs.~\cite{essler-LiebLiniger, rottoli2025eigenstate} is rooted in the structure of thermal charge fluctuations. Our findings also hint at a possible universality: despite the free-boson quantum field theory might at first glance appear microscopically quite different from the Lieb--Liniger and Heisenberg models, all these models nevertheless exhibit the same thermal distribution of matrix elements. We emphasize, however, that statistical properties of matrix elements are not determined solely by the macrostate, but instead depend in general on the choice of the averaging ensemble, and particularly on the scale of charge fluctuations.

\section{General framework}\label{sec:generalframework}

\subsection{Integrability structure of the massless boson field theory}

We describe a class of quantum field theories in $1+1$ dimensions which involve a single scalar bosonic field $\hat{\varphi}(\mathrm{x})$. In a compactified space with circumference $L$, the field admits the Fourier mode expansion,
\begin{equation}
\hat{\varphi}(\mathrm{x}) = - i \sum_{n\neq 0} \frac{\hat{a}_n}{n} e^{- i (2 \pi n/L)\mathrm{x}},
\end{equation}
with mode operators $\hat{a}_n$ satisfying the commutation relations of Heisenberg algebra $[\hat{a}_n, \hat{a}_m ] = n\,\delta_{n,-m}$ with $\hat{a}_n^{\dagger} = \hat{a}_{-n}$.~\footnote{To be more precise, we consider only the holomorphic (right-moving) part of the scalar field due to decoupling of the two chiral components, as is usually done in CFTs. In addition, we drop the zero-mode $\hat{a}_0$ and the position operator $\hat{q}_{0}$ from the expansion of $\hat{\varphi}(x)$ as they have no impact on objects of our study.}
The field can be associated a $U(1)$ current
\begin{equation}\label{U(1)-current}
\hat{J}(\mathrm{x})= - \partial_{\mathrm{x}} \hat{\varphi}(\mathrm{x}) = \frac{2 \pi}{L} \sum_{ n \neq 0} \hat{a}_n e^{-i (2 \pi n/L)\mathrm{x}},
\end{equation}
which is used to construct an infinite one-parameter family of commuting charges, $\hat{Q}^{(k)}=\int^{L}_{0}{\rm d}\mathrm{x}\, \hat{q}^{(k)}(\mathrm{x})$, depending on the coupling parameter $\beta \in \mathbb{R}$. For instance, the densities of the first two charges in the hierarchy read

\begin{equation}\label{qBOcharges}
\begin{aligned}
    \hat{q}^{(1)} &= \frac{1}{4 \pi} :\hat{J}^{2}:, \\
    \hat{q}^{(2)} &= \frac{1}{4 \pi} \left( \frac{1}{3}:\hat{J}^{3}: - \frac{1}{2}(\beta - \beta^{-1}) : \partial_{\mathrm{x}}\hat{J} ~ \mathbb{H}[\hat{J}]: \right),
\end{aligned}
\end{equation}
where $\mathbb{H}[f](\mathrm{x}) =\frac{1}{L} \text{p.v.} \int_{0}^{L} \dd \mathrm{y} \text{ }f(\mathrm{y}) \text{ ctg}\left( \frac{\pi}{L} (\mathrm{x} - \mathrm{y} ) \right) $ denotes the Hilbert transform for $L$-periodic functions and $:\bullet:$ indicates the normal ordering of $\hat{a}_n$. The simplest charge $\hat{Q}^{(1)}$ is easily recognized as the Hamiltonian of a massless free boson. We nevertheless emphasize that $\hat{Q}^{(1)}$ on its own does not suffice to uniquely fix a hierarchy of commuting charges due to the freedom in choosing the coupling parameter $\beta$; to select a particular model one thus needs to fix the value of $\beta$. To phrase it differently, the simplest Hamiltonian $\hat{Q}^{(1)}$ appears as a member of infinitely many distinct hierarchies of charges that do not commute between each other. This somewhat peculiar property is not limited to the massless free boson, but rather is a distinct feature of all 2D conformal field theories.

The hierarchy of charges \eqref{qBOcharges} -- widely known in the literature as the quantum Benjamin--Ono (qBO) hierarchy -- has been extensively studied in Ref.~\cite{Nazarov_2013}. The name reflects the fact that upon the rescaling $\hat{J} \to \beta^{-2} \hat{v}$ and taking the limit $\beta \to 0$, the charges become the conserved quantities of the classical Benjamin--Ono equation~\cite{KAUP1998123} for a classical scalar field $v$.

We specialize subsequently to the $\beta=1$ case which, as we shortly discuss now, leads to certain simplifications. The key observation in this regard is that the structure of eigenstates and matrix elements depends on $\beta$ in a very regular fashion, so we expect no qualitative changes of our results for other values of $\beta$.
Also notice that the `pseudolocal' character of the qBO charges stemming from the Hilbert transform in Eq.~\eqref{qBOcharges} is lost at $\beta=1$ where strict locality is restored. More importantly, for arbitrary values of $\beta$, there exist a class of operators whose matrix elements are explicitly known. But at $\beta=1$ these operators are precisely the vertex operators of the free boson CFT, representing all local primary fields of the theory.

Also, at $\beta=1$ the theory admits an equivalent fermionic formulation \cite{Gavrylenko_2016}. However, we find it advantageous to work with the bosonic representation to take full advantage of the Young diagram representation. In addition to that, the bosonic language permits introducing fluctuations at the level of macrostate in a rather straighforward fashion.

\paragraph*{Spectrum at $\beta=1$.}
Setting the coupling to $\beta=1$, it is instructive to have a look
at the initial two charges in the Fourier mode representation, 
\begin{equation}\label{charges}
\begin{aligned}
    \hat{Q}^{(1)} & = \left( \frac{2 \pi}{L} \right) \sum_{n>0} \hat{a}_{-n} \hat{a}_{n}, \\
    \hat{Q}^{(2)} &=  \frac{1}{6} \left( \frac{2 \pi}{L} \right)^2 \!\!\!  \sum_{i+j+k = 0} :\hat{a}_i \hat{a}_j \hat{a}_k:.
\end{aligned}
\end{equation}
The eigenvectors of these operators are constructed using Schur functions $s_{\boldsymbol{{\lambda}}}(x_1, x_2, ...)$ labeled by partitions $\boldsymbol{\lambda} = ( \lambda_1 , \lambda_2 , ... , \lambda_l)$ subject to $\lambda_1 \geq \lambda_2  \geq  ... \geq \lambda_l > 0$ \cite{macdonald1995}.
Using the power-sum symmetric functions $p_k = \sum_i x_i^{k}$, we can write them as
\begin{equation}\label{Schur-poly}
\begin{aligned}
    s_{(1)} &= p_1, \\
    s_{(2)} &= \frac{1}{2} ( p_1^2 + p_2), \\
    s_{(1,1)} &= \frac{1}{2} (p_1^2 - p_2),
\end{aligned}
\end{equation}
and so forth. The eigenvectors are then constructed by acting with $s_{\boldsymbol{\lambda}}$ on the vacuum state $\ket{\varnothing}$, defined by the relations $\hat{a}_{n} \ket{\varnothing} = 0$ for all $n>0$, while using the replacement rule $p_k \rightarrow \hat{a}_{-k}$. Accordingly, the previous three polynomials yield the following normalized eigenvectors:
\begin{equation}\label{Schur-eigenstates}
\begin{aligned}
    \ket{(1)} &= \hat{a}_{-1} \ket{\varnothing}, \\
    \ket{(2)} &= \frac{1}{2} (\hat{a}_{-1}^2 + \hat{a}_{-2})\ket{\varnothing}, \\
    \ket{(1,1)} &= \frac{1}{2} (\hat{a}_{-1}^2 - \hat{a}_{-2})\ket{\varnothing}.
\end{aligned}
\end{equation}
It is easy to verify that these states are eigenvectors of the charges~\eqref{charges}. 

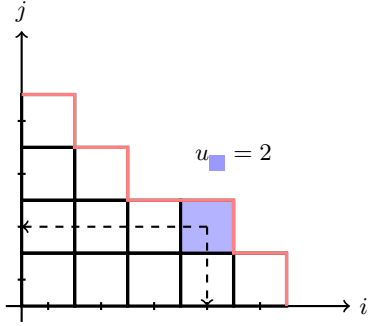
\begin{figure}[H]

\begin{center}
\begin{tikzpicture}[scale=0.7]

\newcommand{\partition}{5,4,2,1}

\def\maxi{4.5}
\def\maxj{3.5}

\foreach \length [count=\row from 0] in \partition {
    \pgfmathtruncatemacro{\n}{\length}
    \foreach \col in {0,...,\numexpr\n-1} {
        \ifnum\col=3
            \ifnum\row=1
                \fill[blue!30] (\col,\row) rectangle ++(1,1);
            \fi
        \fi
        \draw[very thick] (\col,\row) rectangle ++(1,1);
    }
}

\draw[dashed,->, thick] (3.5,1.5) -- (3.5,0); 
\draw[dashed,-> ,thick] (3.5,1.5) -- (0,1.5); 

\node at (4,2.8) { $u_{\colorbox{blue!40}{\phantom{}}} = 2 $ };

\draw[->, thick] (-0.3, 0) -- (6.2, 0) node[right] {$i$};
\draw[->, thick] (0, -0.3) -- (0, 5.2) node[above] {$j$};

\foreach \x in {0,...,\maxi} {
    \draw[thick] (\x+0.5, 0.075) -- ++(0,-0.15);
}

\foreach \y in {0,...,\maxj} {
    \draw[thick] (0.075,\y+0.5) -- ++(-0.15,0);
}

\draw[very thick,color=red!50] (0, 4) -- (1, 4) -- (1,3) -- (2,3) -- (2,2) -- (4,2) -- (4,1) -- (5,1) -- (5,0);

\end{tikzpicture}
\end{center}

\caption{{\bf Eigenstates as Young diagrams.} An example of an eigenstate with $N=12$ excitations encoded by the Young diagram corresponding to partition $\boldsymbol{\lambda} = (5, 4 ,2 ,1)$ with $|\boldsymbol{\lambda}|=N$ boxes. Each cell of $\boldsymbol{\lambda}$, located at an integer-valued coordinate $(i,j)$, carries a Bethe root. For example, the cell marked in blue at position $(4,2)$ is assigned the Bethe root $u_{\colorbox{blue!40}{\phantom{}}} = 4-2=2 $. The discrete curve depicted in red color is a discrete shape function $\psi_{\boldsymbol{\lambda}}(x)$ of this diagram, see Eq.~(\ref{discrete-shape}). }

\label{Young-and-Betheroots}

\end{figure}

To each eigenvector one can assign a unique set of Bethe roots. To do so, it is useful to represent a partition as a Young diagram, i.e. as a set of identical cells (boxes) positioned at coordinates $\{ ( i ,j) \in \mathbb{N} \times \mathbb{N} | \text{ }1\leq j \leq l , \text{ } 1\leq i \leq \lambda_j   \}$. A cell $\square$ with coordinates $( i_{\square},j_{\square})$ is assigned the Bethe root $u_{\square} = i_{\square} - j_{\square}$, see Fig.~\ref{Young-and-Betheroots}. Eigenvalues of $\hat{Q}^{(k)}$ are then given by the power sums of the Bethe roots summed over the whole Young diagram\footnote{The spectrum of qBO systems with a generic value of coupling $\beta$ can be constructed in an analogous fashion, but instead using the Jack polynomials and Bethe roots of the form $u_{\square} =  \beta^{-1} (i_{\square}-1) - \beta (j_{\square} -1)$, see \cite{Alba_2011} or \cite{BelavinBelavin}.}:
\begin{equation}\label{eigenvalues}
    \hat{Q}^{(k)} \ket{\boldsymbol{\lambda}} = \left(\frac{2\pi}{L}\right)^k \sum_{\square \in \boldsymbol{\lambda}} u_{\square}^{k-1} \ket{\boldsymbol{\lambda}}.
\end{equation}
The special case of $k=1$ is understood with the prescription $0^0 \rightarrow 1$, so the eigenvalues of $\hat{Q}^{(1)}$ are simply proportional to the number of cells in the Young diagram.

In summary, we are considering a model that features an infinite family of conserved charges \eqref{charges} built from the current of a massless bosonic field that can be diagonalized by entirely combinatorial means.
In particular, eigenvectors can be described in terms of the Schur functions \eqref{Schur-poly}, \eqref{Schur-eigenstates}, while charge eigenvalues are additive expressions involving the Bethe roots \eqref{eigenvalues}, which are in simple correspondence with the cells of the Young diagrams, see Fig.~\ref{Young-and-Betheroots}.

\subsection{Distance between eigenstates and macrostates description}\label{section-macrostates}

A diagram $\boldsymbol{\lambda}$ can be represented by a discrete \emph{shape function} as
\begin{equation}\label{discrete-shape}
    \psi_{\boldsymbol{\lambda}} (x) = \text{ number of } \text{ }  j \in \mathbb{N} \quad \text{ s.t. } \quad \lambda_j >x ,
\end{equation}
corresponding to a discrete curve along the zig-zag border of the diagram as depicted in Fig.~\ref{Young-and-Betheroots}. Discrete shape functions offer a simple way to measure the distance between eigenstates:
the distance between two states $\boldsymbol{\lambda}$ and $\boldsymbol{\mu}$ can, for instance, be the number of boxes that are not shared by the two diagrams 
$\text{dist}(\boldsymbol{\lambda} , \boldsymbol{\mu})  \sim | \boldsymbol{\mu}/(\boldsymbol{\mu} \cap \boldsymbol{\lambda}) | + | \boldsymbol{\lambda} / (\boldsymbol{\mu}\cap \boldsymbol{\lambda}) |$.
Using the $L^{1}$-norm on the space of discrete shape functions, we choose the following distance:
\begin{equation}\label{distance-Young}
    \text{dist} ( {\boldsymbol{\lambda} ,\boldsymbol{\mu}} ) \equiv \frac{1}{L} \int_{0}^{\infty} | \psi_{\boldsymbol{\lambda}} - \psi_{\boldsymbol{\mu}} | \dd x. 
\end{equation}
We included the rescaling factor $1/L$ here to guarantee that this distance upper bounds charge fluctuations $\delta Q^{(k)}$ for any $k$. Moreover, in the ensembles of states we will consider the scaling relation $\text{dist}(\boldsymbol{\lambda} , \boldsymbol{\mu}) = O(L^{\gamma})$ matches exactly with the magnitude of charge fluctuations $\delta Q^{(k)} = O(L^{\gamma})$.

\paragraph*{Macrostates and limit shapes.}
We now consider the thermodynamic limit of Young diagrams by taking  the system size large while ensuring the charge values scale extensively, i.e. $Q^{(k)}=O(L)$.
It follows from Eq.~\eqref{eigenvalues} that this requires $\sum_{\square} u_{\square}^{k} = O(L^{k+2})$, which is automatically satisfied when the number of cells scales proportionally to $L^{2}$. Therefore, both cell coordinates $(i_{\square}, j_{\square})$ must be rescaled linearly with $L$, and thus $u_{\square} \sim L$. At the same time, the sum over all cells contributes an extra factor of $L^{2}$, and therefore the states $\boldsymbol{\lambda}$ with extensively growing charges are precisely those whose number of cells grows proportionally to $L^2$.
By suitably rescaling both coordinates $(i,j)$ with $L$, we introduce the rescaled shape function
\begin{equation}\label{rescaled-shape}
\psi_{\boldsymbol{\lambda}}^{L} (x) = \frac{1}{L} \psi_{\boldsymbol{\lambda}} (Lx).
\end{equation}
The ensemble of microstates belongs to a particular macrostate provided that asymptotically
\begin{equation}
\psi_{\boldsymbol{\lambda}}^{L} (x) = \psi(x) + o(1).    
\end{equation}
Here $\psi(x)$ is the \emph{limit shape} function that uniquely encodes a macrostate. It follows from its definition that $\psi'(x)<0$ and $\psi(x)>0$ for all $x$, as well as $\psi(+\infty) = 0$. For technical reasons we will assume $\psi(x)$ to be at least twice differentiable.

In the macrostate given by $\psi(x)$, the densities $q^{(k)}$ of local charges $Q^{(k)} = q^{(k)}L + o(L)$ read explicitly

\begin{equation}\label{charges-f}
    q^{(k)} = (2\pi)^k \int_{0}^{\infty}\dd x \int_{0}^{\psi(x)}\dd y\,(x-y)^{k-1}.
\end{equation}

\paragraph*{Macrostate entropy.}
Macrostates carry a combinatorial entropy given by the logarithm of the number of microstate realizations. Computing the entropy boils down to counting the total number of distinct discrete shape function with the same limit shape. The entropy density $s$ of the limit shape $\psi(x)$ takes the form~\footnote{Upon changing the variable to the Bose field occupation function $n(x)=-\psi^{\prime}(x)$ this is none other than the Bose--Einstein statistical weight. However, in terms of $n(x)$ the charges do not admit a simple integral representation akin to Eq.\eqref{charges-f}.}
(see Appendix~\ref{entropy-derivation} for details)
\begin{equation}\label{entropy-f}
    s[\psi']= \int_{0}^{\infty} \left[ (1-\psi') \text{log}(1-\psi') + \psi'\text{log} (-\psi') \right] \dd x,
\end{equation}
where $\psi' \equiv \dd\psi(x)/\dd x$ (note that $\psi' \leq 0$).

\paragraph*{Free energy.}
We now describe how to determine the macrostates of (generalized) Gibbs
ensembles
\begin{equation}
    \hat{\rho}_{\boldsymbol{\upmu}} = Z^{-1}\,e^{-\sum_n \upmu_n \hat{Q}^{(n)}},  
\end{equation}
characterized by chemical potentials $\boldsymbol{\upmu}=\{\upmu_{n}\}_{n=1}^{\infty}$. 

At large $L$, the corresponding partition sum $Z = \sum_{\boldsymbol{\lambda}} e^{-\sum_{n} \upmu_n Q_{\boldsymbol{\lambda}}^{(n)}}$ can be represented via the functional integral
\begin{equation}
    Z = \int D [\psi(x)] e^{-L \mathcal{F}_{\boldsymbol{\upmu}}[\psi,\psi']},
\end{equation}
with the Lagrangian
\begin{equation}
    \mathcal{F}_{\boldsymbol{\upmu}}[ \psi,\psi'] = \sum_{n} \upmu_n q^{(n)}[\psi] - s[\psi'].   
\end{equation}
The dominant contribution to this integral comes from the saddle-point limit shape $\psi_{*}(x)$ that satisfies $\delta \mathcal{F}|_{\psi=\psi_{*}} = 0$, and therefore $Z= e^{-L  \mathcal{F}_{\boldsymbol{\upmu}}}|_{\psi=\psi_{*}} $. Using Eqs.~\eqref{charges-f} and \eqref{entropy-f}, the saddle point equation can be written as
\begin{equation}\label{limit-equation}
    \psi'' = \psi' (\psi' - 1) \sum_{n=1}^{\infty} (2\pi)^n \upmu_n (x-\psi)^{n-1},
\end{equation}
with the prescribed boundary conditions $\psi(+\infty) = 0$, $\psi(0) = +\infty$ (see Appendix \ref{entropy-derivation} for a derivation). Equation \eqref{limit-equation} takes the role of the thermodynamic Bethe ansatz equations for macrostates in the space of Young diagrams.

\paragraph*{Vershik ensemble.}
We explicitly consider the simplest Gibbs ensemble with $\upmu_1 \equiv \upmu$ and  $\upmu_{n\geq2} = 0$, corresponding to the canonical ensemble on the set of all Young diagrams whose Boltzmann weights are given by the number of cells.
This ensemble has previously been considered in the literature by Vershik in~\cite{vershik-stat}, and the corresponding limit shape is often referred to as the \emph{Vershik curve}. The later can be reobtained within our approach by solving Eq.\eqref{limit-equation}, yielding
\begin{equation}\label{Vershik-ls}
    \psi_V (x) = - \frac{1}{2 \pi \upmu} \text{log} \left( 1  - e^{-2\pi \upmu x} \right).
\end{equation}
The number of boxes under the curve equals $N_V = L^{2} \int_{0}^{\infty} \psi(x)dx = L^2 / (24 \upmu^2)$, while the entropy density is $s_V = \pi / (6 \upmu)$.

\subsection{Vertex operators}

The vertex operators $\hat{V}_{\alpha}( \rm{x}) = :e^{i \alpha \hat{\varphi}( \rm{x} )}:$ represent a distinguished class of local observables in our theory. In conjunction with the associated current $\hat{J}(\rm{x})$, vertex operators constitute the complete set of primary fields of the free-boson CFT with central charge $c = 1$. Other fields in the theory are thus conformal descendants and can be accessed in a systematic way using the general CFT methods.

We subsequently focus on the vertex operators and (with no loss of generality) set the coordinate $\rm{x}$ to zero, i.e. we consider $\hat{V}_{\alpha} \equiv \hat{V}_{\alpha}(0)$ which in terms of Fourier mode operators factorizes as
\begin{equation}\label{vertex}
    \hat{V}_{\alpha} = \text{exp}\left( -\alpha \sum_{n>0} \frac{\hat{a}_{-n}}{n} \right) \text{exp} \left( \alpha \sum_{n>0} \frac{\hat{a}_n}{n} \right).
\end{equation}

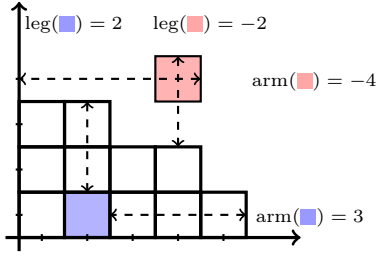
\begin{figure}[htb]

\begin{center}
\begin{tikzpicture}[scale=0.6]

\newcommand{\partition}{5,4,2}

\def\maxi{4.5}
\def\maxj{3.5}

\foreach \length [count=\row from 0] in \partition {
    \pgfmathtruncatemacro{\n}{\length}
    \foreach \col in {0,...,\numexpr\n-1} {
        \ifnum\col=1
            \ifnum\row=0
                \fill[blue!30] (\col,\row) rectangle ++(1,1);
            \fi
        \fi
        \draw[very thick] (\col,\row) rectangle ++(1,1);
    }
}

\fill[red!30] (3,3) rectangle ++(1,1);
\draw[thick] (3,3) rectangle ++(1,1);
\draw[<->,dashed, thick] (3.5,4) -- (3.5,2); 
\draw[<->,dashed, thick] (4,3.5) -- (0,3.5); 

\draw[dashed,<->, thick] (1.5,3) -- (1.5,1); 
\draw[dashed,<->, thick] (5,0.5) -- (2,0.5); 

\draw[->, very thick] (-0.3, 0) -- (6.2, 0); 
\draw[->, very thick] (0, -0.3) -- (0, 5.2);

\foreach \x in {0,...,\maxi} {
    \draw[thick] (\x+0.5, 0.075) -- ++(0,-0.15);
}

\foreach \y in {0,...,\maxj} {
    \draw[thick] (0.075,\y+0.5) -- ++(-0.15,0);
}





\node at (6.4,0.45) { \scriptsize $\text{arm} ( {\raisebox{0.6ex}{\colorbox{blue!40}{\phantom{}}}} ) = 3$ };

\node at (1.2,4.7) { \scriptsize $\text{leg} ( {\raisebox{0.6ex}{\colorbox{blue!40}{\phantom{}}}} ) = 2$ };

\node at (6.5,3.45) { \scriptsize $\text{arm} ( {\raisebox{0.6ex}{\colorbox{red!35}{\phantom{}}}} ) = -4$ };

\node at (4.2,4.7) { \scriptsize $\text{leg}( {\raisebox{0.6ex}{\colorbox{red!35}{\phantom{}}}} ) = -2$ };

\end{tikzpicture}
\end{center}

\caption{{\bf Introducing arms and legs.} Each cell of $\boldsymbol{\lambda}$ is assigned the \emph{arm} and \emph{leg} length as exemplified by the two cells in the depicted diagram $\boldsymbol{\lambda}=(5,4,2)$ marked by blue and red colors, respectively.
}

\label{arms-legs}

\end{figure}

\paragraph*{Schur basis.}
The aim is now to compute the matrix elements of $\hat{V}_{\alpha}$ for a general pair of states from the Fock space. To accomplish this task, one can use the commutation relation $[\hat{a}_n , \hat{V}_{\alpha}] = - \alpha \hat{V}_{\alpha}$ that holds for all $n$, together with $\bra{\varnothing} \hat{V}_{\alpha} \ket{\varnothing} = 1$. When expanded in an arbitrary Fock space basis, the matrix elements are given by certain polynomials in $\alpha$ whose roots reveal no obvious pattern. The remarkable property of the Schur basis, see Eqs.~\eqref{Schur-poly} and \eqref{Schur-eigenstates}, is that the roots of these polynomials are intimately linked with combinatorial properties of the corresponding Young diagrams. In particular, matrix elements can be expressed using the so-called arms and legs ascribed to the cells as follows: given a diagram $\boldsymbol{\lambda}$, each cell $\square$ is assigned the \emph{arm}, $a_{\boldsymbol{\lambda}}(\square)\equiv \text{arm}_{\boldsymbol{\lambda}}(\square)$, as the number of cells in $\boldsymbol{\lambda}$ located to the right of cell $\square$. Similarly,
the \emph{leg} of a cell, $l_{\boldsymbol{\lambda}}(\square)\equiv\text{leg}_{\boldsymbol{\lambda}} (\square)$, is the number of cells in $\boldsymbol{\lambda}$ above $\square$. For example, in the Fig.~\ref{arms-legs} the arm and the leg of the blue cell are $3$ and $2$, respectively\footnote{While we draw Young diagrams using the so-called French convention (with rows increasing from bottom to top), we borrow our `arm' and `leg' terminology from the English convention (which adopts the top-down row ordering). In the English convention, the cell referred to as the `head', together with its arm and leg cells, resembles a human figure, hence the origin of these terms.}. Alternatively, one can imagine the arm (or leg) as a horizontal (or vertical) distance between the right (or top) side of the cell and the discrete curve marking the outer border of the diagram. This formulation allows to extend the definitions of arms and legs also to the cells that are not a part of the diagram $\boldsymbol{\lambda}$; in this case, however, we need to define them with the extra minus sign. As exemplified in Fig.~\ref{arms-legs}, the arm and leg of the red cell are $-4$ and $-2$, respectively. Finally, we introduce the \emph{hooks} via
$h_{\boldsymbol{\lambda}}\equiv {\rm hook}_{\boldsymbol{\lambda}}(\square)={\rm leg}_{\boldsymbol{\lambda}}(\square)+{\rm arm}_{\boldsymbol{\lambda}}(\square)+1$.

Using the above conventions, and by introducing $ a^{(\alpha)}_{\boldsymbol{\mu}\boldsymbol{\lambda}}(\square)\equiv a_{\boldsymbol{\mu}}(\square)-a_{\boldsymbol{\lambda}}(\square)+\alpha$,
the matrix elements of $\hat{V}_{\alpha}$ of eigenstates $\boldsymbol{\lambda}$ and $\boldsymbol{\mu}$ in the Schur-basis representation admit a compact form:
\begin{equation}\label{matrix-element}
    \bra{\boldsymbol{\lambda}} \hat{V}_{\alpha} \ket{\boldsymbol{\mu}} = 
    \prod_{\square \in \boldsymbol{\lambda}} \left[ 1 + \frac{ a^{(\alpha)}_{\boldsymbol{\mu}\boldsymbol{\lambda}}}{h_{\boldsymbol{\lambda}}} \right]
    \prod_{\square \in \boldsymbol{\mu}} \left[ 1 - \frac{a^{(\alpha)}_{\boldsymbol{\mu} \boldsymbol{\lambda}}}{h_{\boldsymbol{\mu}}} \right].
\end{equation}




We note that, despite complete factorization over cells, the above expression is not simply a product over two independent terms involving individual diagrams $\boldsymbol{\lambda}$ and $\boldsymbol{\mu}$ due to their mutual `interaction' contained in $a^{(\alpha)}_{\boldsymbol{\mu}\boldsymbol{\lambda}}$.

Formula \eqref{matrix-element} can be easily verified explicitly for the first few states using Eq.~\eqref{Schur-eigenstates}. To establish it rigorously it proves helpful to employ the mapping between the Schur states and fermionic representation by means of bosonization~\cite{Gavrylenko_2016}. It is instructive to mention here that similar formulae for matrix elements arise in a more general setting. Indeed, analogous formulae have been discovered for the entire Benjamin--Ono hierarchy \cite{Nazarov_2013} upon an appropriate modification of the vertex operator. Moreover, similar formulae have been derived in Ref.~\cite{Alba_2011} for the ${\rm qBO_2}$ case.

Factorizable matrix elements (\ref{matrix-element}) of vertex operators, Eq.~\eqref{vertex}, is our key technical tool for studying statistical properties of matrix elements. We stress, however, that in spite of their simplicity, the products over all cells render the computation still highly non-trivial in the relevant regime of large $L$. What further complicates the asymptotic analysis are spurious singularities that appear at the diagram boundaries upon substituting the hooks $h_{\boldsymbol{\lambda}}$ with the leading terms in their large-$L$ expansion. As we demonstrate in the next section, it is nevertheless still possible to carry out both analytical and numerical investigations in an efficient manner.

\section{Scaling properties of matrix elements }
\label{sec:matr_elem}

We are now ready to finally investigate the structure of the off-diagonal matrix elements of vertex operators $\hat{V}_{\alpha}$ in the thermodynamic limit. We will particularly consider their logarithmic squared modulus,
\begin{equation}
    \kappa_{ \boldsymbol{\lambda}  \boldsymbol{\mu} } (\alpha) = -\frac{1}{L}\log|\bra{\boldsymbol{\lambda}}\hat{V}_{\alpha} \ket{\boldsymbol{\mu}}|^2,
\end{equation}
with eigenstates $\boldsymbol{\lambda}$ and $\boldsymbol{\mu}$ sampled from a prescribed statistical ensemble. Our main task will be to analyze their distributions, i.e. probability density functions of
$\kappa_{ \boldsymbol{\lambda}  \boldsymbol{\mu} }$, and to characterize the scaling properties of their mean $\kappa = \langle \kappa_{\boldsymbol{\lambda} \boldsymbol{\mu}} \rangle$ and width $\delta \kappa = \sqrt{\langle \kappa_{\boldsymbol{\lambda} \boldsymbol{\mu}}^2 \rangle - \langle \kappa_{\boldsymbol{\lambda} \boldsymbol{\mu}} \rangle^2}$.

Before delving to this analysis we begin with a fully analytic study of the states comprising the so-called \emph{staircase diagrams}. Here, owing to their simple structure of arms and legs, we succeeded in deriving an exact asymptotic formula for the suppression rate depending as a function of distance \eqref{distance-Young} between two such diagrams. Exploring this toy example turned out to be tremendously useful for revealing certain universal features of the factorizable formula for $\bra{\boldsymbol{\lambda}}V_{\alpha} \ket{\boldsymbol{\mu}}$, which we subsequently support in a more general setting with our numerical study of matrix elements in modified Vershik ensembles.

\subsection{Analytical results: staircase diagrams}\label{sec:staircasediagrams}

In the following, we give an analytical calculation for the class of staircase diagrams, namely partitions of type $\boldsymbol{\lambda}_k = (k, k-1, ..., 2, 1)$ with $k \in \mathbb{N}$.

We consider a pair of diagrams $\boldsymbol{\lambda}_L$ and $\boldsymbol{\lambda}_{L+d}$ with separation $d \in \mathbb{N}$, whilst assuming $L \in \mathbb{N}$ (otherwise
it suffices picking a diagram with $k= \lfloor L \rfloor$). According to distance \eqref{distance-Young}, we have
$\text{dist}(\boldsymbol{\lambda}_L , \boldsymbol{\lambda}_{L+d}) = d + \frac{d(d-1)}{2L} $ which, as long as $d \ll L$, is approximately $\text{dist}(\boldsymbol{\lambda}_L , \boldsymbol{\lambda}_{L+d}) \approx d$ for large $L$. If we furthermore impose the scaling $d\sim L^{\gamma}$ with $\gamma<1$, both diagrams approach the same macrostate characterized by the limit shape $\psi_0(x) = 1 - x $ with $x<1$. On the other hand, putting $d = \mathrm{q}\,L$, the two diagrams flow towards different macrostates, one with the limit shape $\psi_0(x) = 1-x$ for $x\leq1$ and the other $\psi_\mathrm{q}(x) = 1+\mathrm{q} - x$ for $x\leq1+\mathrm{q}$.

To each pair of staircase diagrams separated by a distance $d$ we associate a function
\begin{equation}
    \kappa(L,d;\alpha) \equiv \kappa_{\boldsymbol{\lambda}_L \boldsymbol{\lambda}_{L+d}}(\alpha),
\end{equation}
which we subsequently analyze in the limit $L \rightarrow \infty$ for different scaling exponents $\gamma$, namely $d \sim L^{\gamma}$.

As a warmup, we consider first the case of $d=0$, i.e. matrix elements with $\boldsymbol{\mu}=\boldsymbol{\lambda}$ \footnote{We purposely refrain from calling such matrix elements diagonal for the following reason: upon reinstating the zero mode $\hat{a}_0$, alongside the associated position operator $\hat{q}$, the eigenstates of the model acquire (besides a Young diagram $\boldsymbol{\lambda}$) an additional zero-mode quantum number $\hat{P}$, defined via $\hat{a}_0 \ket{\boldsymbol{\lambda} , P} = \hat{P} \ket{\boldsymbol{\lambda} ,P}$. Matrix elements of the vertex operator $\hat{V}_{\alpha}$ should in this case instead of Eq.~\eqref{matrix-element} be written as $\bra{\boldsymbol{\lambda} , P'} V_{\alpha} \ket{\boldsymbol{\mu} , P} \propto \delta_{P'-P, \alpha}$. Therefore, even by picking $\boldsymbol{\lambda} = \boldsymbol{\mu}$, $V_{\alpha}$ in reality always couples eigenstates with the difference of the zero-mode quantum numbers equal to $\alpha$}. In this case Eq.~\eqref{matrix-element} simplifies as
\begin{equation}
    \bra{ \boldsymbol{\lambda}} \hat{V}_{\alpha} \ket{\boldsymbol{\lambda}} = \prod_{\square \in \boldsymbol{\lambda}} \left( 1 - \frac{\alpha^2}{h^2_{\boldsymbol{\lambda}}(\square)} \right). 
\end{equation}
For staircase diagrams, the number of cells in $\boldsymbol{\lambda}_{L}$ with hooks of a given length equals $N_{\square}(h_{\boldsymbol{\lambda}_L}=2k+1) = L-k$, $N_{\square}(h_{\boldsymbol{\lambda}_L}=2k) = 0$, for $0 \leq k\leq L-1$, yielding
\begin{equation}
    \bra{ \boldsymbol{\lambda}_L} \hat{V}_{\alpha} \ket{ \boldsymbol{\lambda}_L} = \prod_{k=0}^{L-1} \left(1 - \frac{\alpha^2}{(2k+1)^2} \right)^{L-k}.
\end{equation}
This product admits the following large-$L$ asymptotics
\begin{equation}
    \kappa(L,0;\alpha) = -\log\left( 
\cos^{2}(\tfrac{\pi}{2}\alpha) \right) + o(1).
\end{equation}
Evidently, for any value of $\alpha \in \mathbb{R}\setminus \mathbb{Z}$~\footnote{For $\alpha \in \mathbb{Z}$, vertex operators $\hat{V}_{\alpha}$ become sparse in the chosen basis, compatibly with a singular behavior of $\kappa(L,0;\alpha)$ at these values.}, $\kappa(L,0;\alpha)=O(1)$, implying exponential decay at large $L$ (with a non-trivial rate depending on $\alpha$).

We next turn to the general case of $d>0$, which, despite being slightly more complicated, is still tractable. In particular, for $ d \sim L^{\gamma}$ with $\gamma<1$, the behavior of $\kappa(L,d;\alpha)$ in the large-$L$ limit is captured by the asymptotic formula (see Appendix~\ref{app-triangle-mel} for details)
\begin{equation}\label{lines-larged}
    \kappa = \kappa_{0}(d;\alpha) + \frac{u}{2} + O(L^{2\gamma -1}),
\end{equation}
depending on the scaling variable
\begin{equation}\label{u-variable}
    u = \frac{d^2}{L}\log{\left(\frac{L}{d}\right)},
\end{equation}
and a constant $\kappa_{0}(d;\alpha)=- \log \left[\cos^{2} \left(\tfrac{\pi}{2}(\alpha+d)\right) \right]$.
This formula proves tremendously useful since it encodes information about the rate of suppression for all fluctuation scales $\gamma$ at once. By plugging in $d = O(L^{\gamma})$, one can readily see that the first term stays of the order $O(1)$, whereas the second term produces a correction of the form $L^{2 \gamma -1} \log{(L)}$.

Lastly, we analyze the case of different macrostates by setting $d = \mathrm{q}\,L$. At the leading order in $L$:
\begin{equation}
    \kappa(L,\mathrm{q}L;\alpha) = k(\mathrm{q}) L + o(L),
\end{equation}
where $k(\mathrm{q})$ is given by Eq.~\eqref{eq:kdeflin}.

We conclude by succinctly gathering our conclusions regarding the suppression rate of the staircase diagrams depending on the distance $d = O(L^{\gamma})$ for $0\leq \gamma\leq 1$:
\begin{enumerate}
    \item For $\gamma < 1/2$, the first term in Eq.~\eqref{lines-larged} dominates, signifying exponential suppression of the matrix elements with the system size, $\kappa = O(1)$, or equivalently
\begin{equation} 
    |\!\bra{\boldsymbol{\lambda}}\hat{V}_{\alpha}\ket{\boldsymbol{\mu}}|^{2} = e^{-O(L)}.
\end{equation}
    \item In the interval $\gamma \in [1/2 , 1)$, conversely, the second term in Eq.~\eqref{lines-larged} takes over, leading to a stronger than exponential suppression (alongside an extra multiplicative $\log{(L)}$ correction), namely, $\kappa = O(L^{2 \gamma -1} \log L)$, or
    \begin{equation}
        |\!\bra{\boldsymbol{\lambda}}\hat{V}_{\alpha}\ket{\boldsymbol{\mu}}|^{2} = e^{-O(L^{2 \gamma} \log L )}.
    \end{equation}

    \item For $\gamma =1$, corresponding to different macrostates, the suppression gets further enhanced to
    \begin{equation}
        |\!\bra{\boldsymbol{\lambda}}\hat{V}_{\alpha}\ket{\boldsymbol{\mu}}|^{2} = e^{-O(L^{2})}.
    \end{equation}
\end{enumerate}

\subsection{Modified Vershik ensemble}
\label{sec:Mod_Vershik}

In Sec.~\ref{section-macrostates} we introduced the Vershik ensemble as a canonical Gibbs state on the space of Young diagrams with the number of boxes as statistical weights. In addition, we already derived the limit shape of the Vershik ensemble, cf. Eq.~\eqref{Vershik-ls}. 

For definiteness, we here fix the chemical potential to $\upmu=1/\sqrt{24}$, which corresponds to fixing the average number of boxes to $N_V  = L^2$. The limit shape curve then reads explicitly
\begin{equation}\label{limit_Versh}
    \psi_V(x) = - \frac{\sqrt{6}}{\pi} \text{log} \left( 1 - e^{-(\pi/\sqrt{6})x} \right),
\end{equation}
with the corresponding entropy density $s_V = \pi \sqrt{2/3}$.

Besides the exact limit shape, the fluctuations in the Vershik ensemble have also been characterized explicitly; they are governed by a Gaussian process with a magnitude of $O(L^{1/2})$ that can be described in terms of a `fluctuation field'
\begin{equation}\label{def_g}
g_{\boldsymbol{\lambda}}(x) = \sqrt{L} \left( \psi_{ \boldsymbol{\lambda} }^{L} (x) - \psi_V (x)  \right).
\end{equation}
Here $\psi_{\boldsymbol{\lambda} }^{L} (x)$ is the \emph{rescaled} shape function~\eqref{rescaled-shape} of the diagram sampled from the Vershik ensemble. At large $L$, this fluctuation field $g_{ \boldsymbol{\lambda} } (x) $ converges to a Gaussian process $g_V(x)$ with zero mean and covariance $\mathbb{E}[g_V (x) g_V (y)] =: C_V(x,y)$ of the form \cite{FreimanVershikYakubovich1997,Yakubovich2001,Funaki_2013}
\begin{equation}\label{eq:norm_cov_matrix}
     C_V(x,y) = - \frac{\sqrt{6}}{\pi} \psi'_V \left( \text{max}[x,y] \right).
\end{equation}

Thus, asymptotically in $L$, sampling states from the Vershik ensemble can be implemented by sampling the integer sequences $\{\lambda_j'\}$,
\begin{equation}\label{eq:norm_sampling}
    \lambda_j' \sim \lfloor L\,\psi_{V} \left( j /L \right) + L^{1/2} \, g_V \left( j / L \right)\rfloor,
\end{equation}
and sorting them in a non-increasing way, $\{\lambda^{\prime}_{j}\}\mapsto \{\lambda_{j}\}$, to obtain valid Young diagrams $\boldsymbol{\lambda}=(\lambda_{1} , \lambda_{2}, \ldots )$.

This procedure permits us to modify the Vershik ensemble by replacing the fluctuation scale $L^{1/2}$ with a general scale $d=O(L^{\gamma})$,
\begin{equation}\label{eq:mod_sampling}
    \lambda_j' \sim \lfloor L\,\psi_{V} \left( j /L \right) + d \cdot \, g_V \left( j / L \right)\rfloor,
\end{equation}
which we refer to as the \emph{modified Vershik ensembles}.

As discussed in Appendix~\ref{sampling-appendix}, the special form of $g_{V}(x)$ with the covariance matrix~\eqref{eq:norm_cov_matrix} enables us to generate samples of states for system sizes of the order $L\simeq  10^6$.

The outlined sampling algorithm enables efficient numerical computation of the statistical properties of pseudo-random variables $\kappa_{\boldsymbol{\lambda}\boldsymbol{\mu}}=L^{-1}\log|\bra{\boldsymbol{\lambda}}V_{\alpha}\ket{\boldsymbol{\mu}}|^{2}$. In what follows, we implement the following averaging protocol: both states $\boldsymbol{\lambda}$ and $\boldsymbol{\mu}$ are sampled according to Eq.\eqref{eq:mod_sampling} with $d$ fixed. In such a sampling $d$ is proportional to the mean distance between the sampled states. Note, however, that such ensembles are only well-defined in the regime $1\ll d\ll L$: for small $d$ the rounding effects can lead to severe corrections to the target Gaussian process, whereas for $d\simeq L$ systematic corrections arise from sorting.

\subsection{Numerical results}

In this section, we gather the key results of our numerical analysis of statistical properties of matrix elements $\kappa_{\boldsymbol{\lambda}\boldsymbol{\mu}}$.

\subsubsection{Scaling exponents}
\label{sec:scaling-exp}

\begin{figure}[h!]
 \includegraphics[width=1.0\columnwidth]
 {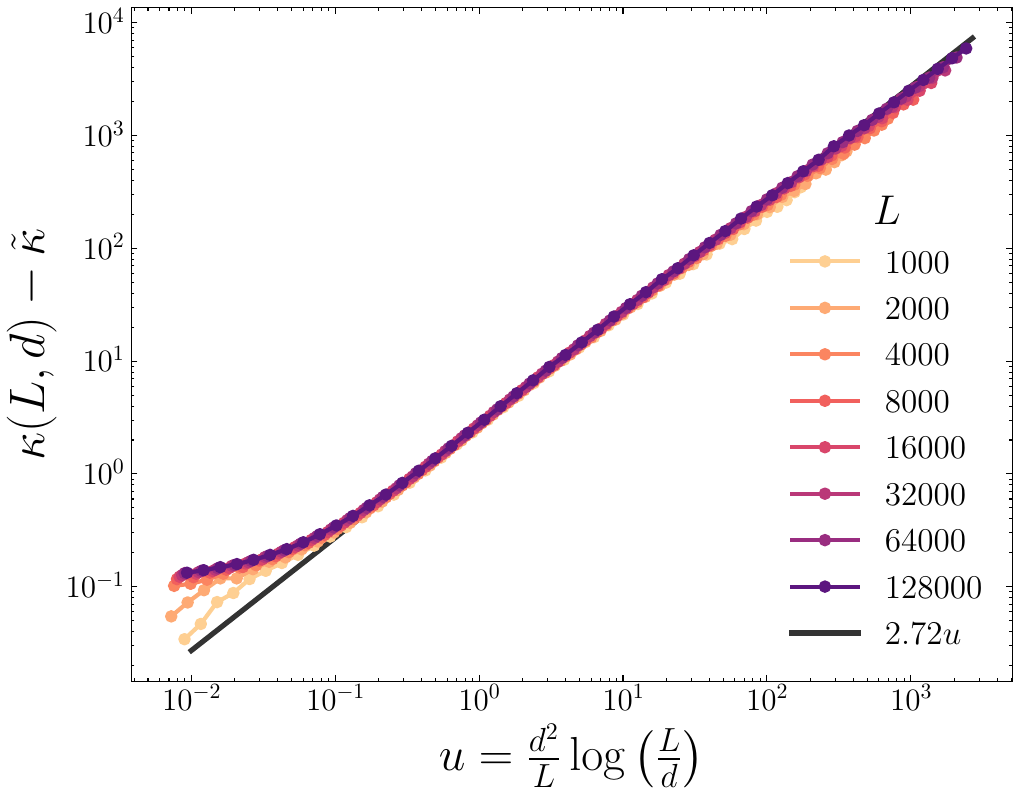}
\caption{{\bf Scaling behavior of suppression rate.} The suppression rate (distribution mean) $\kappa(L,d;\alpha)$ as a function of the scaling variable $u = \tfrac{d^2}{L} \log {(L/d)}$, shown for the vertex operator with $\alpha=0.2$ for various system sizes. Solid black line shows the asymptotic function $\kappa(u)=2.72 u + \tilde{\kappa}$ with $\tilde{\kappa}\approx 3$ obtained from the numerical fit at $L=8 \cdot 10^{3}$ within the interval $u \in (0.1,100)$.}

\label{scaling-Vershik_a}
\end{figure}

Motivated by the analytical result, Eq.~\eqref{lines-larged}, we express the distribution mean $\kappa(L,d)$ in the modified Vershik ensemble in terms of the scaling variable $u = (d^{2}/L)\log(L/d)$.
The data shown in Fig.~\ref{scaling-Vershik_a} indicates that $u$ remains a good scaling variable for $\kappa$:
\begin{equation}\label{eq:scaling_mod_k}
    \kappa=f_{\kappa} \left(u\right).
\end{equation}

As also shown in Fig.~\ref{scaling-Vershik_a}, in the small-$u$ regime, $f_\kappa$ saturates to a finite constant $\kappa_0$ (which, as in the case of staircase diagrams, depends on parameter $\alpha$ of the vertex operator $\hat{V}_{\alpha}$). Note that taking $d = O(L^{\gamma})$ with $0\leq\gamma < 1/2$ corresponds to $u \to 0$ in the large-$L$ limit. Thus, in this regime, the suppression rate remains intensive, $\kappa = \kappa_0 = O(1)$, which implies that the scaling exponent introduced in Eq.~\eqref{scaling-exponents} equals $p(\gamma<1/2)=0$. In contrast, at large $u$, $\kappa$ exhibits linear asymptotics $f_\kappa(u) \sim u$ leading to $\kappa = O(L^{2\gamma-1} \log L)$ for $\gamma \geq 1/2$. As a result, the scaling exponent in this regime is $p(\gamma\geq1/2) = 2 \gamma -1$. Altogether, we can see that the suppression rate $\kappa$ in the modified Vershik ensemble exhibits behavior analogous to the case of staircase diagrams.

\begin{figure}[h!]
 \includegraphics[width=1.0\columnwidth]
 {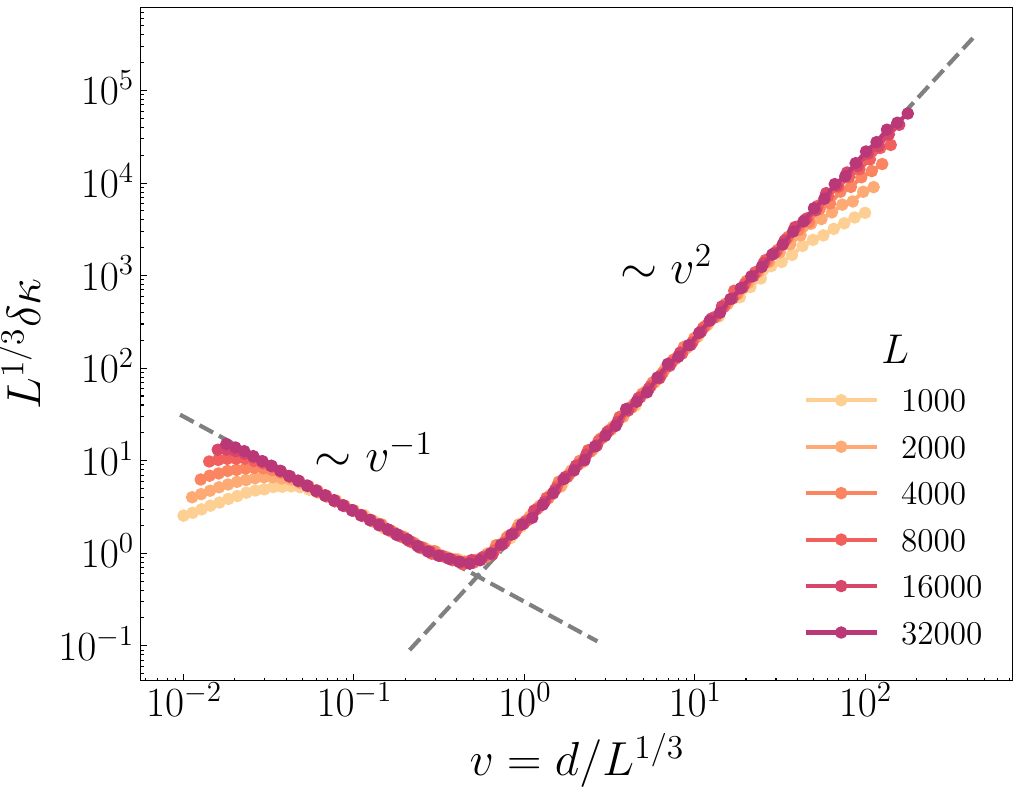}
 \caption{{\bf Scaling behavior of distribution width.} The rescaled distribution width $L^{1/3}\,\delta\kappa(L,d; \alpha)$ as a function of the scaling variable $v = d / L^{1/3}$ for $\alpha = 0.2$ and various system sizes. Dashed lines indicate the asymptotic behaviors described approximately by $0.3\,v^{-1}$ at small $v$ and $2v^2$ at large $v$, respectively.
 }
\label{scaling-Vershik_b}
\end{figure}

Characterizing the scaling behavior of the distribution width $\delta \kappa$ is more intricate, and we empirically found that the distribution width obeys the scaling form
\begin{equation}
    \label{eq:scaling_function}
    \delta \kappa = L^{-1/3} f_{\delta\kappa}(d/L^{1/3}).
\end{equation}
As illustrated in Fig.~\ref{scaling-Vershik_b}, this scaling collapse is well satisfied, with the scaling function obeying $f_{\delta \kappa}(v) \sim v^{-1}$ for small $v$ and $f_{\delta \kappa}(v) \sim v^{2}$ for large $v$. In particular, these asymptotics imply the scaling exponents $q(\gamma \leq 1/3) = -\gamma$ and $q(\gamma > 1/3) = 2\gamma - 1$.

In Appendix \ref{sec:scaling_analysis} we perform a systematic scaling analysis to confirm that $\kappa$ and $\delta\kappa$ satisfy the scaling laws given by Eqs. \eqref{eq:scaling_function} and \eqref{eq:scaling_mod_k} in the range $1\ll d \ll L$.

We also note that the scaling function $f_{\delta \kappa}(v)$ exhibits a sharp crossover 
joining two distinct asymptotic regimes at $v_{\star} \approx 0.45$, see Fig.~\ref{scaling-Vershik_b}. This crossover becomes particularly relevant for fluctuation scales $\gamma < 1/3$, where 
it is manifested as the system-size dependence of $\delta \kappa$. The 
asymptotic scaling $\delta \kappa = O(L^{-\gamma})$ can be observed only when 
$d / L^{1/3} \lesssim v_{\star}$, whereas for $d / L^{1/3} \gtrsim v_{\star}$ one 
instead finds the apparent scaling $\delta \kappa = O(L^{2\gamma - 1})$.

Finally, we also examined the scaling exponents using an alternative 
sampling procedure. Instead of the “perfect” discretization of the limit shape 
$\psi_{V}$ used previously, see Eq.~\eqref{eq:mod_sampling}, can first sample a seed configuration 
$\boldsymbol{\lambda}_{\mathrm{seed}}$ from the Vershik ensemble,  
Eq.~\eqref{eq:norm_sampling}, and then generate pairs of states in its 
$d$-vicinity following the same procedure, supplemented by an additional averaging over the seed states.
Although this approach exhibits slower convergence with system size, it 
yields consistent results for the scaling exponent $p(\gamma)$ across the 
full range of scales $\gamma$. For $q(\gamma)$, reliable extraction is possible 
only for $\gamma \geq 1/2$, where our conclusions are again consistent with those found using our primary sampling scheme Eq.~\eqref{eq:mod_sampling} (see Appendix \ref{sec:scaling_analysis}).

\subsubsection{Distributions of matrix elements}\label{sec:distributions}

Having analyzed the scaling exponents of the suppression rate and distribution width, we now examine the full probability density function 
$P(\mathcal{K}_{\boldsymbol{\lambda}\boldsymbol{\mu}}) \equiv \mathrm{PDF}(\mathcal{K}_{\boldsymbol{\lambda}\boldsymbol{\mu}})$ of matrix elements parametrized by the normalized pseudorandom variable
\begin{equation} \label{preudo_def}
 \mathcal{K}_{\boldsymbol{\lambda}\boldsymbol{\mu}} \equiv \frac{\kappa_{\boldsymbol{\lambda}\boldsymbol{\mu}} - \kappa}
    {\delta \kappa}.
\end{equation}

\begin{figure}[H]
\includegraphics[width=1.0\linewidth]{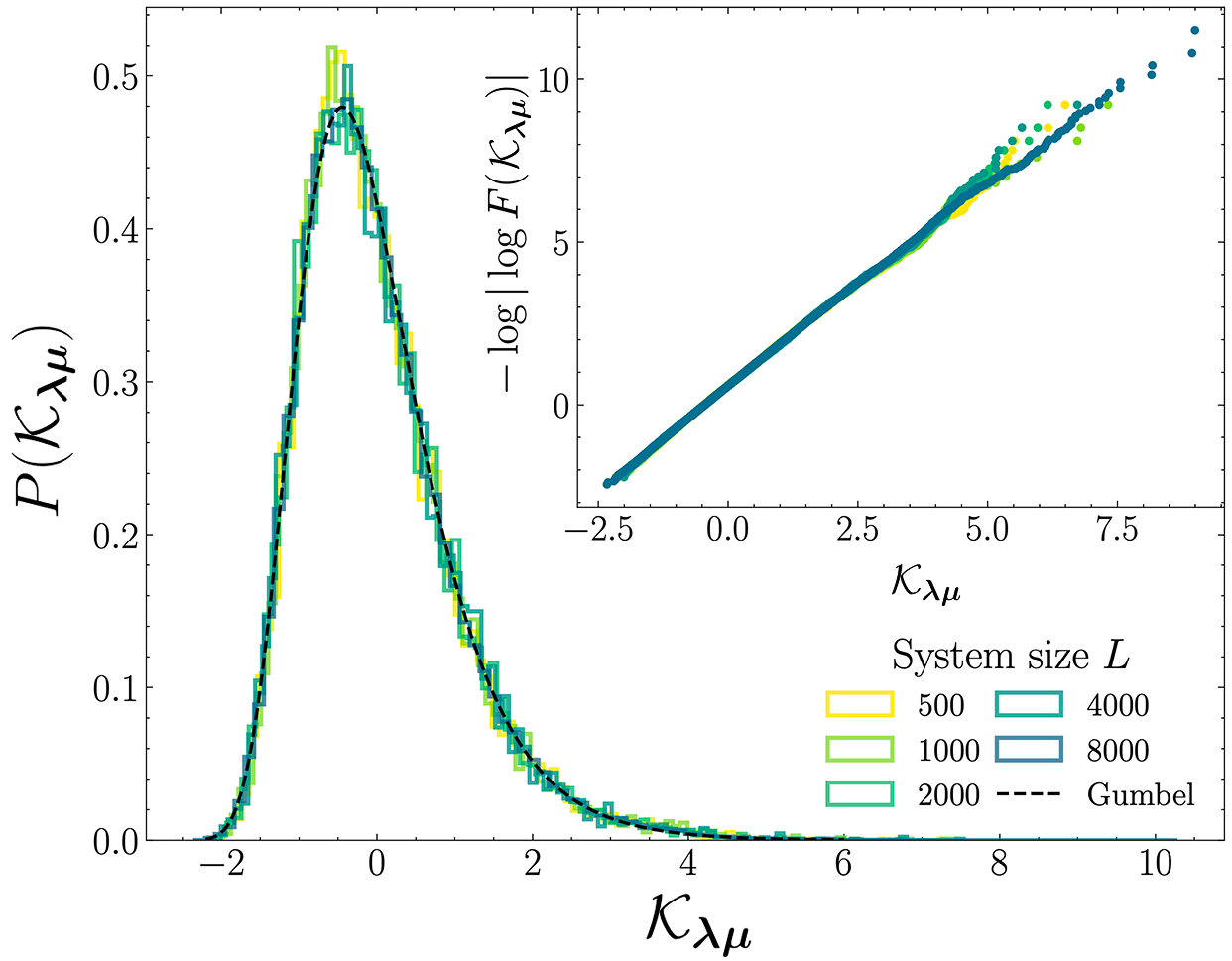}
\centering
\caption{{\bf Probability distribution of matrix elements in the thermal ensemble.} (main panel) PDF of normalized matrix elements $\mathcal{K}_{\boldsymbol{\lambda}\boldsymbol{\mu}}$ drawn in the thermal (Vershik) ensemble computed for various system sizes. Black line shows the fit with the Gumbel distribution. (inset) the doubly-logarithmic form of ${\rm CDF}(\mathcal{K}_{\boldsymbol{\lambda}\boldsymbol{\mu}})$ and the linear prediction of the Gumbel CDF \eqref{CDF}.}
\label{fig:Gumb_L}
\end{figure}

To facilitate a direct comparison with the thermal Gibbs ensembles employed in Refs.~\cite{essler-LiebLiniger, rottoli2025eigenstate}, we first examine the Vershik ensemble with the fluctuation scale $\gamma = 1/2$. The data, shown in Fig.~\ref{fig:Gumb_L}, shows a nearly perfect collapse with the Gumbel distribution $P_{\rm G}(\mathcal{K})$, with the cumulative density function (CDF) $F_{\rm G}(\mathcal{K}) \equiv \int_{-\infty}^{\mathcal{K}} P_{\rm G}(\mathcal{K}')\, d \mathcal{K}'$ of the form
\begin{equation}
    F_{\rm G}(\mathcal{K}) = \exp{\big(-\exp{(-(\mathcal{K} - \mu_{\rm G})/\beta_{\rm G}})\big)}.\label{CDF}
\end{equation}
Here $\beta_{\rm G} = \sqrt{6}/\pi$ and $\mu_{\rm G} = -\sqrt{6}\upgamma/\pi$ are fixed by normalization \eqref{preudo_def}, where $\upgamma$ is the Euler constant. This prediction is in line with the numerical analysis of Ref. \cite{rottoli2025eigenstate} done for the Heisenberg spin chain.

\begin{figure}[htb]
\includegraphics[width=1.0\linewidth]{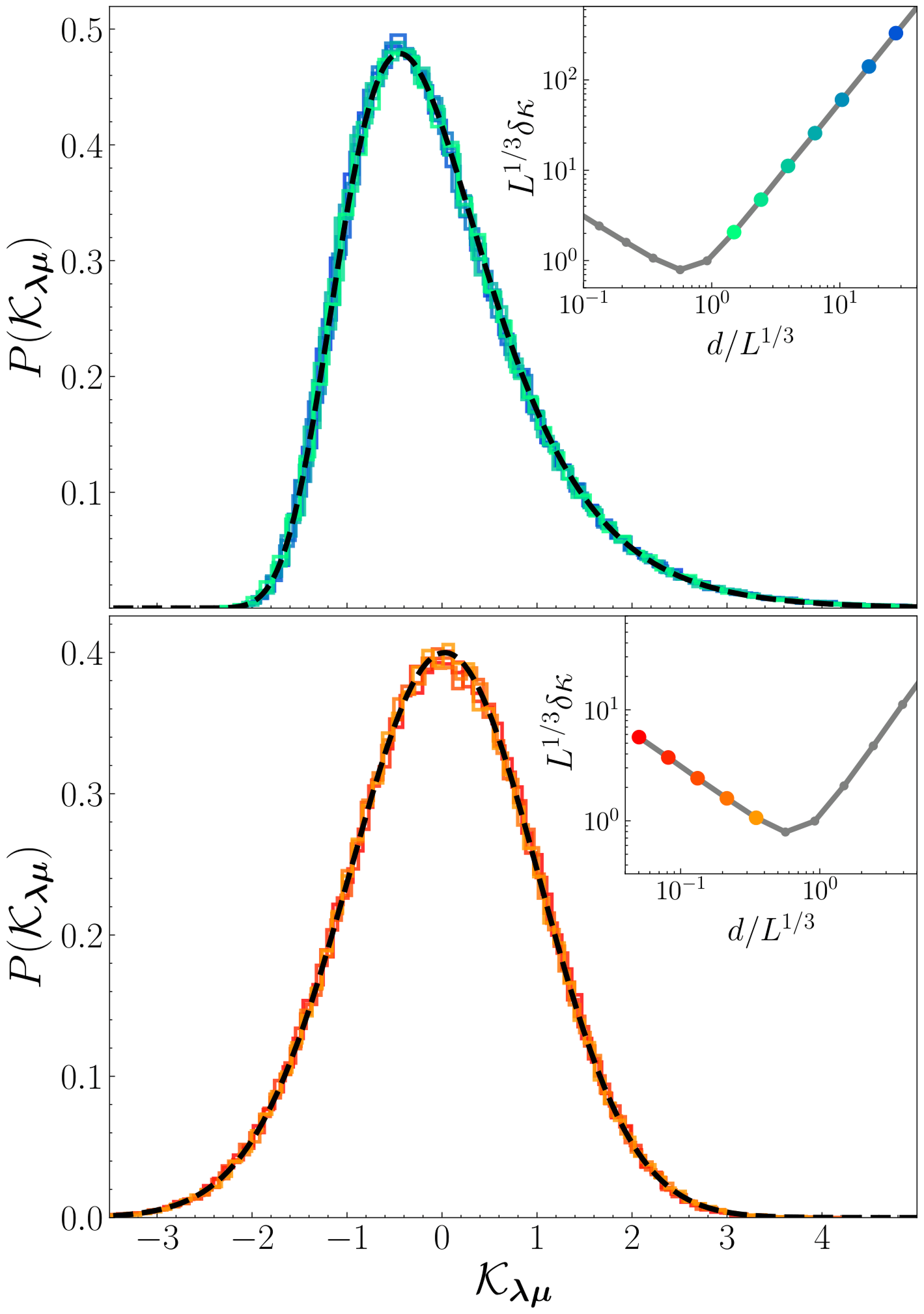}
\centering
\caption{{\bf Probability distribution of matrix elements in the modified Vershik ensembles.} PDF of $\mathcal{K}_{\boldsymbol{\lambda}\boldsymbol{\mu}}$ in the modified Vershik ensemble shown for different values of the scaling variable $v=d/L^{1/3}$ at system size $L=16\cdot 10^{3}$ in the regimes $v > v_{\star}$ (upper panel) and $v<v_{\star}$ (lower panel). Black lines show the fitting distributions: the Gumbel distribution (upper panel) and the skew normal distribution with parameter $\varepsilon \approx 0.6812$ (lower panel). The insets show the the scaling function $f_{\delta \kappa}(v)$ with representative points marked by different colors.}
\label{fig:Two_dist}
\end{figure}

We go further and investigate the fluctuation-scale dependence of the distributions in the regime $0<\gamma<1$ for the modified Vershik ensembles. In this regard, it is crucial to pay attention to two distinct asymptotic regimes of the scaling function $f_{\delta \kappa}$ which is responsible for a crossover phenomenon at a critical fluctuation scale $\gamma_{\star}=1/3$. We thus consider separately the regimes $\gamma<\gamma_{\star}$ and $\gamma>\gamma_{\star}$ using the common scaling variable $v=d/L^{1/3}$.
As shown in Fig.~\ref{fig:Two_dist}, we find two distinct probability distributions $P(\mathcal{K}_{\boldsymbol{\lambda}\boldsymbol{\mu}})$:

\begin{itemize}
    \item $v>v_{\star}$: above the critical fluctuation scale $\gamma>\gamma_{\star}$, including the thermal scale $\gamma=1/2$, $\mathcal{K}_{\boldsymbol{\lambda}\boldsymbol{\mu}}$ are distributed according to the Gumbel distribution with CDF \eqref{CDF}.
    
    \item $v<v_{\star}$: in this regime $\mathcal{K}_{\boldsymbol{\lambda}\boldsymbol{\mu}}$ follow an asymptotic PDF differing from the Gumbel one, which can be accurately approximated by the skew normal distribution $P_{\rm SN}(\mathcal{K})$ with a skewness parameter $\varepsilon$, which in terms of $\eta_{\varepsilon}(\mathcal{K})\equiv (\mathcal{K}-\mu_{\rm SN})/\beta_{\rm SN}$ reads
 \begin{equation}
 P_{\rm SN}(\mathcal{K};\varepsilon)=
    \frac{e^{-\eta^{2}_{\varepsilon}/2}}{\pi \beta_{\rm SN}}
    \int\limits_{-\infty}^{\varepsilon \eta_{\varepsilon}}e^{-t^2/2} \mathrm{d}t,
\end{equation}
with parameters $\beta_{\rm SN} = \left(1-\frac{2 \varepsilon^2}{\pi(1+\varepsilon^2)}\right)^{-1/2}$ and $\mu_{\rm SN}=-\left(\frac{2\varepsilon^2}{\pi \beta_{\rm SN}(1+\varepsilon^2)} \right)^{1/2}$. 
Note that for $\varepsilon=0$ one recovers the standard normal distribution.
\end{itemize}

\section{Conclusion}
\label{sec:conclusion}

In this work, we outlined a general approach for investigating statistical properties of matrix elements of local observables in quantum many-body systems
which permits to systematically explore their structure on different fluctuation scales.

By introducing a suitable distance between the eigenstates, we implemented our approach in an integrable quantum field theory of a massless boson which enables a purely combinatorial description of the spectrum and, more importantly, admits a compact factorizable formula for the vertex-operator matrix elements amenable to analytic and numerical study.

Our main results are (i) an exact expression for the suppression rate of matrix elements as a function of the eigenstate distance for a particular class of microstates; (ii) an intricate, non-analytic dependence of algebraic scaling exponents (characterizing the suppression rate and distribution width) on the ensemble fluctuation scale; (iii) an extensive numerical analysis of distributions of matrix elements which unveils two distinct regimes, each described by its own universal probability distribution.

Our findings open up several interesting directions for future exploration.
An important task would be to implement a similar analysis for other integrable many-body systems. In addition, despite computational limitations, attempting a similar approach in chaotic systems may offer valuable information about the structure of local observables at large energy separations. 

Although our present approach relies heavily on numerical sampling, we believe  that statistical properties of matrix elements in the studied model might be accessible by entirely analytic means with the aid of combinatorial methods applied to Young diagrams. Indeed, statistical properties of hook lengths remain an active topic of mathematical research, however, with only limited results currently available~\cite{mutafchiev2019,griffin2022}.

Regarding physics applications, it is imperative to understand if or how our statistical characterization of matrix elements facilitates the computation of the corresponding dynamical correlation functions in integrable systems. Since correlation functions in the eigenbasis representation involve sums over the entire spectrum of eigenstates, it is essential to identify class of matrix elements are most weakly suppressed and represent the relevant contribution to the spectral sum in the thermodynamic limit~\cite{senese2025}. This step requires a careful consideration as distributions of matrix elements themselves cannot be simply restored from the leading asymptotics of the distributions describing the suppression rates. 


\section*{Acknowledgments}
The authors thank F. Essler and L. Vidmar for insightful discussions, and A. Dymarsky, F. Fritzsch, T. Prosen, R. Senese for fruitful comments on the manuscript.
EI is supported by the Research Program P1-0402 and Project N1-0368 funded by
the Slovenian Research Agency (ARIS). R.S. acknowledges support from ERC Advanced
Grant No. 101096208—QUEST, and Research Programme
P1-0402 of Slovenian Research and Innovation Agency
(ARIS).

\appendix

\section{Entropy of the limit shape and TBA equations}\label{entropy-derivation}

Entropy of a macrostate encoded by the limit shape $\psi(x)$ is equal to the logarithm of the total number of Young diagrams $\boldsymbol{\lambda}$ with the discrete shapes $\psi_{\boldsymbol{\lambda}} (x)$, cf. Eq.~\ref{discrete-shape}, that approximate the rescaled limit shape, $ \psi_{\boldsymbol{\lambda}} (x) \approx L \psi(x/L)$. 

To compute it explicitly, we discretize the horizontal axis $x$ into equal segments by introducing $\{ x_{i} \}_{i=0}^{\infty}$ such that $x_{0} = 0$ and $x_{i+1} - x_{i} = \Delta x$ $\forall i$, assuming that $ 1 \ll \Delta x \ll L$. Local changes of the discrete shape function inside each segment $[x_{i} , x_{i+1} ]$ with prescribed left $\psi_{\boldsymbol{\lambda}} (x_i) = L \psi(x_i/L)$ and right $\psi_{\boldsymbol{\lambda}} (x_{i+1}) = L \psi (x_{i+1} / L )$ values yield different Young diagrams from the same macrostate. The total number $N_{[x_{i},x_{i+1}]}$ of such modifications counts the number of discrete paths from the point $\left( x_{i} ,  L \psi(x_i / L) \right)$ to $\left( x_{i+1} ,  L \psi(x_{i+1}/L) \right)$ and thus can be simply written as a binomial coefficient 
\begin{equation}
\begin{aligned}
    &N_{[x_{i},x_{i+1}]} = \binom{[1+ |\psi'(x_{i}/L)|] \Delta x}{\Delta x} \\
    \approx \text{exp} &\left[ \Delta x \left( (1 + |\psi'|)\text{log}(1+|\psi'|) - |\psi'| \text{log} |\psi'| \right) \right].
\end{aligned}
\end{equation}
In the second line we used the Stirling formula $n! \approx e^{n\log{(n)}}$ valid for large $\Delta x$. Finally, the total number of diagrams can be written as a product $N_{\psi} = \prod_{i} N_{[x_{i},x_{i+1}]}$ and entropy density $s[\psi]  = \frac{1}{L}\text{log } N_{\psi}$ is given by
\begin{equation}
\label{eqn:entropy_density}
\begin{aligned}
    &s = \int\limits_{0}^{+\infty}\!\!{\rm d}x \left( (1 - \psi'  )  \log(1-\psi') + \psi' \log|\psi'| \right).
\end{aligned}
\end{equation}

Employing the above expression for the entropy we can now determine the limit shape that corresponds to the (generalized) Gibbs ensembles $\hat{\rho} = Z^{-1}\,e^{-\sum_n \upmu_n \hat{Q}^{(n)}}$. The corresponding partition function $Z={\rm Tr}(\hat{\rho})$ can be evaluated using saddle-point of the
free-energy functional 
\begin{equation}
    \mathcal{F}_{\boldsymbol{\upmu}}[\psi,\psi'] = \sum_{n} \upmu_n q^{(n)}[\psi] - s[\psi'],    
\end{equation}
with the entropy density given by Eq.~\eqref{eqn:entropy_density} and the charge densities reading 
\begin{equation}
    q^{(n)}[\psi] = \frac{(2\pi)^n}{n} \int\limits_{0}^{+ \infty} \dd x [ x^{n} - (x-\psi(x))^{n}].
\end{equation}
The saddle-point condition, $\delta \mathcal{F}|_{\psi_{\star}} = 0$,
determines a unique limit shape function $\psi_{\star}(x)$, representing a monotonically decreasing function from $\mathbb{R}_{+}$ to $\mathbb{R}_{+}$. The requirement that charge densities attain finite values after evaluating them at $\psi_{*}(x)$ constrains the asymptotic behavior of $\psi_{*}(x)$ at $x=0$ and $x=+\infty$. In particular, to guarantee that $q_1 < \infty$ we must impose $\psi(+{\infty}) = 0$.

Expressing the free-energy functional in the form, $\mathcal{F}_{\boldsymbol{\upmu}} = \int_{0}^{+ \infty} \dd x ~\mathrm{f}[x,\psi,\psi']$, we obtain
\begin{equation}\label{variation}
    \delta \mathcal{F}_{\boldsymbol{\upmu}} = \frac{\partial \mathrm{f}}{\partial \psi'} \delta \psi \Bigg|_{0}^{\infty} + \int\limits_{0}^{+\infty} \mathrm{d}x \left[ \frac{\partial \mathrm{f}}{ \partial \psi}  -\frac{\mathrm{d}}{\mathrm{d}x} \left( \frac{\partial \mathrm{f}}{\partial \psi'} \right) \right] \delta\psi.
\end{equation}
Since $\frac{\partial \mathrm{f}}{\partial \psi'} = \log \left( \frac{1-\psi'}{-\psi'} \right)$, the first term can indeed be nullified by imposing the boundary conditions $\psi(0) = +\infty$ and $\quad \psi(+\infty) = 0$, in which case $\psi'(0) = - \infty$, $\psi'(+\infty) = 0$. Meanwhile, requiring the second term to vanish yields the Euler-Lagrange equations,
\begin{equation}
    \frac{\partial \mathrm{f}}{\partial \psi} - \frac{\dd}{\dd x} \left( \frac{\partial \mathrm{f}}{\partial \psi'} \right) = 0,
\end{equation}
reading explicitly
\begin{equation}
    \psi'' = \psi' (\psi'-1) \sum_{n=1}^{+ \infty} (2 \pi)^n (x-\psi)^{n-1}.
\end{equation}
The solution to this differential equation gives the saddle-point shape function $\psi_{\star}(x)$.

\section{Matrix elements for the staircase diagrams}
\label{app-triangle-mel}

We compute $ | \langle \boldsymbol{\lambda}_{L} | \hat{V}_{\alpha} | \boldsymbol{\lambda}_{L+d} \rangle  |^2$ for two staircase diagrams $\boldsymbol{\lambda}_{L} = (L, L-1...,2 , 1)$ and $\boldsymbol{\lambda}_{L+d} = (L+d, L+d-1, ... , 2 , 1)$ by considering separately contributions to \eqref{matrix-element} from three subsets of cells: the cells shared by both diagrams (\textcolor{blue!80}{blue}), the cells belonging to $\boldsymbol{\lambda}_{L+d}$ which lie to the right of $\boldsymbol{\lambda}_{L}$ (\textcolor{red!80}{red}), and the remaining cells (\textcolor{orange!80}{orange}) from $\boldsymbol{\lambda}_{L+d}$ (lying above $\boldsymbol{\lambda}_{L}$).
Individual contribution to $|\langle \boldsymbol{\lambda}_{L} |\hat{V}_{\alpha}| \boldsymbol{\lambda}_{L+d} \rangle |^2$ from each of these pieces will be subsequently denoted by labels $\mathscr{A}$, $\mathscr{B}$ and $ \mathscr{C}$, respectively. 

\begin{figure}[H]
\centering
\begin{tikzpicture}[scale=0.4]

\def\partitionSmall{6,5,4,3,2,1}
\def\partitionLarge{9,8,7,6,5,4,3,2,1}

\colorlet{colorb}{red!15}

\fill[colorb] (8,0) rectangle ++(1,1);
\fill[colorb] (7,0) rectangle ++(1,1);
\fill[colorb] (6,0) rectangle ++(1,1);

\fill[colorb] (7,1) rectangle ++(1,1);
\fill[colorb] (6,1) rectangle ++(1,1);
\fill[colorb] (5,1) rectangle ++(1,1);

\fill[colorb] (6,2) rectangle ++(1,1);
\fill[colorb] (5,2) rectangle ++(1,1);
\fill[colorb] (4,2) rectangle ++(1,1);

\fill[colorb] (5,3) rectangle ++(1,1);
\fill[colorb] (4,3) rectangle ++(1,1);
\fill[colorb] (3,3) rectangle ++(1,1);

\fill[colorb] (4,4) rectangle ++(1,1);
\fill[colorb] (3,4) rectangle ++(1,1);
\fill[colorb] (2,4) rectangle ++(1,1);

\fill[colorb] (3,5) rectangle ++(1,1);
\fill[colorb] (2,5) rectangle ++(1,1);
\fill[colorb] (1,5) rectangle ++(1,1);

\colorlet{colorc}{orange!20}
\fill[colorc] (2,6) rectangle ++(1,1);
\fill[colorc] (1,6) rectangle ++(1,1);
\fill[colorc] (0,6) rectangle ++(1,1);

\fill[colorc] (1,7) rectangle ++(1,1);
\fill[colorc] (0,7) rectangle ++(1,1);
\fill[colorc] (0,8) rectangle ++(1,1);

\colorlet{colora}{blue!10}

\fill[colora] (0,5) rectangle ++(1,1);

\fill[colora] (0,4) rectangle ++(1,1);
\fill[colora] (1,4) rectangle ++(1,1);

\fill[colora] (0,3) rectangle ++(1,1);
\fill[colora] (1,3) rectangle ++(1,1);
\fill[colora] (2,3) rectangle ++(1,1);

\fill[colora] (0,2) rectangle ++(1,1);
\fill[colora] (1,2) rectangle ++(1,1);
\fill[colora] (2,2) rectangle ++(1,1);
\fill[colora] (3,2) rectangle ++(1,1);

\fill[colora] (0,1) rectangle ++(1,1);
\fill[colora] (1,1) rectangle ++(1,1);
\fill[colora] (2,1) rectangle ++(1,1);
\fill[colora] (3,1) rectangle ++(1,1);
\fill[colora] (4,1) rectangle ++(1,1);

\fill[colora] (0,0) rectangle ++(1,1);
\fill[colora] (1,0) rectangle ++(1,1);
\fill[colora] (2,0) rectangle ++(1,1);
\fill[colora] (3,0) rectangle ++(1,1);
\fill[colora] (4,0) rectangle ++(1,1);
\fill[colora] (5,0) rectangle ++(1,1);

\foreach \length [count=\row from 0] in \partitionLarge {
    \pgfmathtruncatemacro{\n}{\length}
    \foreach \col in {0,...,\numexpr\n-1} {
        \draw[semithick] (\col,\row) rectangle ++(1,1);
    }
}

\foreach \length [count=\row from 0] in \partitionSmall {
    \pgfmathtruncatemacro{\n}{\length}
    \foreach \col in {0,...,\numexpr\n-1} {
        \draw[semithick] (\col,\row) rectangle ++(1,1);
    }
}


\draw[very thick, red!60]
(9,0)
-- (9,1)
-- (8,1)
-- (8,2)
-- (7,2)
-- (7,3)
-- (6,3)
-- (6,4)
-- (5,4)
-- (5,5)
-- (4,5)
-- (4,6)
-- (3,6)
-- (3,7)
-- (2,7)
-- (2,8)
-- (1,8)
-- (1,9)
-- (0,9);

\draw[very thick, blue!60]
(6,0)
-- (6,1)
-- (5,1)
-- (5,2)
-- (4,2)
-- (4,3)
-- (3,3)
-- (3,4)
-- (2,4)
-- (2,5)
-- (1,5)
-- (1,6)
-- (0,6);

\draw[->, semithick] (-0.5, 0) -- (10, 0);   
\draw[->, semithick] (0, -0.5) -- (0, 10);   


\def\maxi{8.5}
\def\maxj{8.5}

\foreach \x in {0,...,\maxi} {
    \draw[semithick] (\x+0.5, 0.075) -- ++(0,-0.15);
}

\foreach \y in {0,...,\maxj} {
    \draw[semithick] (0.075,\y+0.5) -- ++(-0.15,0);
}

\node at (-1.3, 2.8) { \scriptsize $L$ };
\draw[<->,semithick] (-0.6,0.5) -- (-0.6, 5.5);

\node at (4.25, -1.3) { \scriptsize $L+d$ };
\draw[<->,semithick] (0.5,-0.6) -- (8.5, -0.6);

\end{tikzpicture}

\end{figure}

Contributions coming from \textcolor{blue!80}{blue} cells can be computed column by column: writing $\mathscr{A} = \prod_{\ell=1}^{L} \Pi(\ell)$, where $\Pi(\ell)$ denotes contributions from the column with $\ell$ cells, and enumerating the cells within such columns from top to bottom by index $j \in \overline{1,\ell}$, we obtain
\begin{equation}
\begin{aligned}
    \Pi (\ell) = \prod_{j=1}^{\ell}\left( 1+ \frac{d+\alpha}{2j-1} \right)^2 \left( 1 - \frac{d+\alpha}{2j-1+2d} \right)^2,
\end{aligned}
\end{equation}
which can in turn be rewritten using the Pochhammer symbol $ [x ]_n \equiv \prod_{i=1}^{n}(x+i-1)$ as
\begin{equation}\label{Pi}
 \Pi(\ell) = \left( \frac{ \left[ \frac{1+d + \alpha}{2} \right]_\ell}{\left[\frac{1}{2}\right]_\ell} \frac{ \left[ \frac{1+d - \alpha}{2} \right]_\ell}{ \left[\frac{1}{2}+d\right]_\ell } \right)^2.
\end{equation}
With aid of asymptotic expansions for the Pochhammer symbol, we now consider the behavior of \eqref{Pi} at large $\ell$. In the regime $\ell,d \rightarrow\infty$ with $d=o(\ell)$ we can write
\begin{equation}
    \log \Pi(\ell) =   2 \text{log}(2)d - \text{log}(2) - \tfrac{1}{2}d^2/\ell + o(1),
\end{equation}
implying $\mathscr{A} = \prod_{\ell=1}^{L} \Pi(\ell)$ satisfies an asymptotic expansion
\begin{equation}
    \log \mathscr{A} = \log (2^{2d-1})L- \tfrac{1}{2}d^2  \log \left(\tfrac{L}{d}\right) + o(L).
\end{equation}
Somewhat counterintuitively, contributions from domain $\mathscr{A}$ at leading order in $L$ actually yield exponential blowup rather decay. This feature suggests, as we will shortly confirm, that such a blow up must be compensated by an equal contribution (of the opposite sign) coming from red cells. With this in mind, we retained the subleading correction to the above asymptotics of $\log \mathscr{A}$.

Another important regime to consider is the scaling regime $d = \mathrm{p} \, \ell$ for large $\ell$ with $\mathrm{p}>0$ where we obtain the asymptotic expression
\begin{equation}
    \log \Pi(\ell) = 2\ell \log\left( \frac{(\mathrm{p}+2)^{\mathrm{p}+2}}{4(\mathrm{p}+1)^{\mathrm{p}+1}} \right) + O(1).
\end{equation}
Putting $d = \mathrm{q}\,L$ and $\ell = \mathrm{a} \, L$ with $\mathrm{a} \in [0,1]$, we can express $\mathrm{p}=\mathrm{q}/\mathrm{a}$ and resum $\Pi(\ell)$
\begin{equation}
\begin{aligned}
\log \mathscr{A}
&= L^2 \int\limits_{0}^{1} \mathrm{d}\mathrm{a} \,
2 \mathrm{a} \log \left[
\frac{\left(\frac{\mathrm{q}+2\mathrm{a}}{\mathrm{a}}\right)^{\frac{\mathrm{q}+2\mathrm{a}}{\mathrm{a}}}}
{4\left(\frac{\mathrm{q}+\mathrm{a}}{\mathrm{a}}\right)^{\frac{\mathrm{q}+\mathrm{a}}{\mathrm{a}}}}
\right]
+ O(L^2) \\
&= \frac{L^2}{2} \log \left(
\frac{(\mathrm{q}+2)^{(\mathrm{q}+2)^2} \, \mathrm{q}^{\mathrm{q}^2}}
{16 (\mathrm{q}+1)^{2 (\mathrm{q}+1)^2}}
\right)
+ O(L^2).
\end{aligned}
\end{equation}

Computing next the contribution from \textcolor{red!80}{red} cells, each row contributes the same amount, that is $\mathscr{B} = \Big[ \prod_{j=1}^{d}\left( 1 - \frac{d+\alpha}{2j-1}  \right)^2 \Big]^{L}$, yielding
\begin{equation}
     \mathscr{B} = \left( \frac{ \left[\frac{1-d-\alpha}{2})\right]_d}{\left[\frac{1}{2}\right]_d} \right)^{2L}.
\end{equation}
From here we readily deduce the following asymptotic form that is valid for both $d=o(L)$ and $d=O(L)$,
\begin{equation}
    \log \mathscr{B} = -\log (2^{2d-1})L -\kappa_{0}(d;\alpha)L + o(L),
\end{equation}
where
\begin{equation}
    \kappa_{0}(d;\alpha) \equiv -\log \left(\cos^2\left[\tfrac{\pi}{2}(\alpha+d)\right] \right).
\end{equation}

Contributions from \textcolor{orange!80}{orange} cells can be likewise computed as a product over rows (with $\mathrm{p} \in \overline{1,d}$ labeling row lengths), $\mathscr{C} = \prod_{\mathrm{p}=1}^{d} \prod_{j=1}^{\mathrm{p}} \left( 1 - \frac{\mathrm{p}+\alpha}{2j-1} \right)^2$, yielding
\begin{equation}
    \mathscr{C} = \left( \prod_{\mathrm{p}=1}^{d} \frac{ \left[\frac{1-\mathrm{p}-\alpha}{2}\right]_{\mathrm{p}}}{ \left[\frac{1}{2}\right]_{\mathrm{p}}} \right)^2,
\end{equation}
with the asymptotic form
\begin{equation}
    \log \mathscr{C} = - \log(2) d^2 + o(d^2).
\end{equation}

Combining all contributions we find that for $d,L \rightarrow\infty$ and $d=o(L)$ the suppression of matrix elements, $\kappa(L,d;\alpha) \equiv - \frac{1}{L} \log | \langle \boldsymbol{\lambda}_{L} | \hat{V}_{\alpha} | \boldsymbol{\lambda}_{L+d} \rangle |^2$, can be written as
\begin{equation}
    \kappa(L,d ; \alpha) =  \kappa_{0}(d;\alpha)+\frac{1}{2}\frac{d^2}{L} \log \left(\frac{L}{d}\right)+o\left(\frac{d^2}{L}\right).
\end{equation}

Meanwhile, for $d=\mathrm{q}\, L$ we have
\begin{equation}
    \kappa(L,\mathrm{q}L;\alpha) = k(\mathrm{q})\,L + o(L),
\end{equation}
with the prefactor
\begin{equation}
    k(\mathrm{q}) = \frac{1}{2}\log \left(\frac{4^{(\mathrm{q}+1)^2+1}(\mathrm{q}+1)^{2(\mathrm{q}+1)^2}}{(\mathrm{q}+2)^{(\mathrm{q}+2)^2} \mathrm{q}^{\mathrm{q}^2}}\right).
    \label{eq:kdeflin}
\end{equation}

\section{Sampling the modified Vershik ensemble}\label{sampling-appendix}
Here we detail the algorithm used in the sampling of eigenstates from the Vershik ensembles introduced in Sec.~\ref{sec:Mod_Vershik}.

The goal is to efficiently sample a random Gaussian fluctuation field $g(x)$ defined in Eq.~\eqref{def_g}, with the covariance matrix of the form
\begin{equation}
    C_{V}(x,y) \equiv \mathbb{E}\left[g(x)g(y)\right]= f_V(\max[x,y]).
\end{equation}
where we introduced $~f_V(x)\equiv -\frac{\sqrt{6}}{\pi} \psi_V'(x)$ with $\psi_V(x)$ denoting the limit shape of the Vershik ensemble given by Eq.~\eqref{limit_Versh}. The key property of this Gaussian process is its Markovianity, 
\begin{equation}
    C_{V}( x, y)=C_{V}( y, y)\qquad \forall \quad x<y.    
\end{equation}
Since the new field values depend only on the previous ones, the sampling complexity can be reduced from $O(N^3)$ (for a generic covariance matrix) to $O(N)$.

To generate a valid Young diagram from the fluctuation field, we sample the above process on a discrete set of points $x$. To this end, we set $x_i=i/L$ with $i=1,2,...,M$, where $M$ represents a cutoff which is adjusted to properly capture the asymptotic decay of $f_V(x)$.

The inductive algorithm for sampling $g(x_i)$ reads as follows:
\begin{itemize}
    \item \textit{Base case} $i=1$. The first point is drawn independently from a normal distribution,
    \begin{equation}
        g(x_1)=\mathcal{N}(0,f_V(x_1)),
    \end{equation}
    with variance $f_V(x_1)$.
    \item \textit{Inductive step}. Assuming the correct covariance structure for all the previous points $j<i$, the next point is updated according to the rule:
    \begin{equation}
        g(x_i)=\mu_{i}+\sigma_i \,\mathcal{N}(0,1),
    \end{equation}
    where $\mu_i=g(x_{i-1})\frac{f_V(x_i)}{f_V(x_{i-1})}$ and $\sigma_i^2=f_V(x_i)-\frac{f_V(x_i)^2}{f_V(x_{i-1})}$.
\end{itemize}
It can be proven by induction that this conditional update rule yields the desired covariance for the fluctuation field $g(x_i)$.

\section{Scaling analysis}
\label{sec:scaling_analysis}

To corroborate the deduced scaling laws, Eqs. \eqref{eq:scaling_mod_k} and \eqref{eq:scaling_function},  here we present the results of a numerical scaling analysis for the case of modified Vershik ensembles.

\paragraph*{Suppression rate.}
Regarding the scaling behavior of the suppression rate (distribution mean) $\kappa(L,d)$, we argued that $\kappa$ depends on a single scaling variable $u=\frac{d^2}{L} \log\left(\frac{L}{d}\right)$, with asymptotics $\kappa(u)\sim u$ at large $u$ and constant value $\kappa(0)>0$.
Our objective here is to verify whether $u$ provides a good scaling variable for all accessible fluctuation scales $\gamma$. In this respect, note that, by plugging in $d=O(L^{\gamma})$, variable $u$ exhibits nontrivial scaling with $L$. To overcome this issue, we introduce here a suitable scaling variable that scales \emph{intensively} with $L$, namely $v_{\gamma}\equiv d/L^{\gamma}$, and perform the scaling analysis for the entire interval $0\leq\gamma\leq 1$. The task boils down to establish the existence of scaling functions
\begin{equation}
    \label{eq:scaling_function_kappa}
   \Phi^{\kappa}_{\gamma}(v_{\gamma}) \equiv \lim_{L\to \infty} \frac{\kappa(L,v_{\gamma} L^{\gamma})}{ L^{2\gamma-1} \log \left( L^{1-\gamma}/v_{\gamma}\right)}. 
\end{equation}

\begin{figure}[H]
\includegraphics[width=1.0\linewidth]{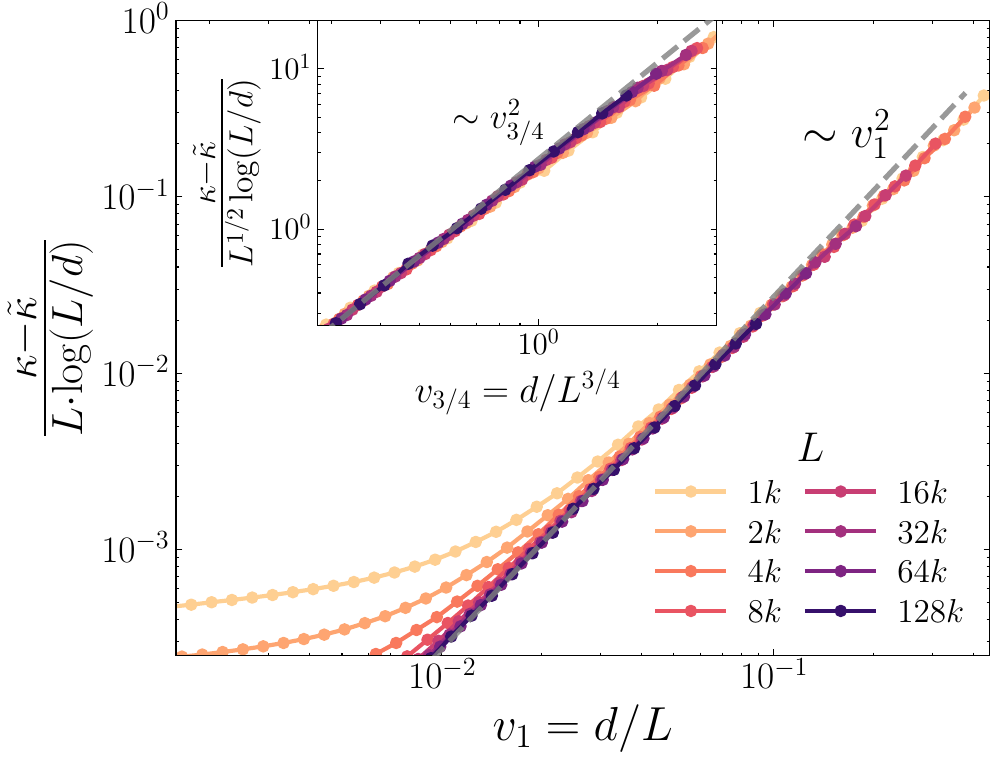}
\centering
\caption{{\bf Scaling analysis of suppression rate $\kappa$.} Convergence towards the asymptotic scaling function $\Phi^{\kappa}_{\gamma}(v_{\gamma})$, cf. Eq.~\eqref{eq:scaling_function_kappa} (with an extra subleading constant shift $\tilde{\kappa} \approx 3$ to achieve better collapse), depending on scaling variable $v_{\gamma}=d/L^{\gamma}$, shown for fluctuation scales $\gamma=1$ (main panel) and $\gamma=3/4$ (inset) for various system sizes. Gray dashed lines represent linear fitting function.}
\label{fig:rescaling_k}
\end{figure}

Since for any $\gamma<1/2$ the rate $\kappa(u)$ with $d=O(L^{\gamma})$ tends to a constant value at large $L$, we only need to scan the interval $\gamma\in [1/2,1)$ where $\kappa(u) \sim u$ for large $L$. In the latter regime, the conjectured asymptotics at large $v_{\gamma}$ reads $\Phi_\gamma^\kappa(v_\gamma)\sim v_\gamma^2$. We found that such asymptotic scaling indeed takes place in the whole interval $1/2 \leq\gamma<1$ by probing the data collapse as it is exemplified for a value of $\gamma=3/4$ in Fig.~\ref{fig:rescaling_k}. We note that an analogous collapse is observed in Fig.~\ref{scaling-Vershik_a} upon plotting $\kappa(L,v_{1/2}L^{1/2})/\log{(L^{1/2})}$ as a function of $v_{1/2}$ (not shown).
At the maximal scale $\gamma=1$ we find the scaling function $\Phi^{\kappa}_{1}(v_{1})$ to exhibit systematic deviations from the quadratic form in the regime $v_{1} \gg 1$.

\paragraph*{Distribution width.}
Regarding the scaling properties of the distribution width $\delta \kappa(L,d)$, we here validate that the conjectured scaling law $\delta \kappa = L^{-1/3} f(d/L^{1/3})$ holds in the regime $1\ll d\ll L$.
This time we introduce the scaling functions
\begin{equation}
    \label{eq:scaling_function_delta_kappa}
    \Phi^{\delta \kappa}_{\gamma}(v_\gamma) \equiv \frac{\delta\kappa(L,v_\gamma L^\gamma)}{L^{q(\gamma)}}.
\end{equation}

\begin{figure}[H]
\includegraphics[width=1.0\linewidth]{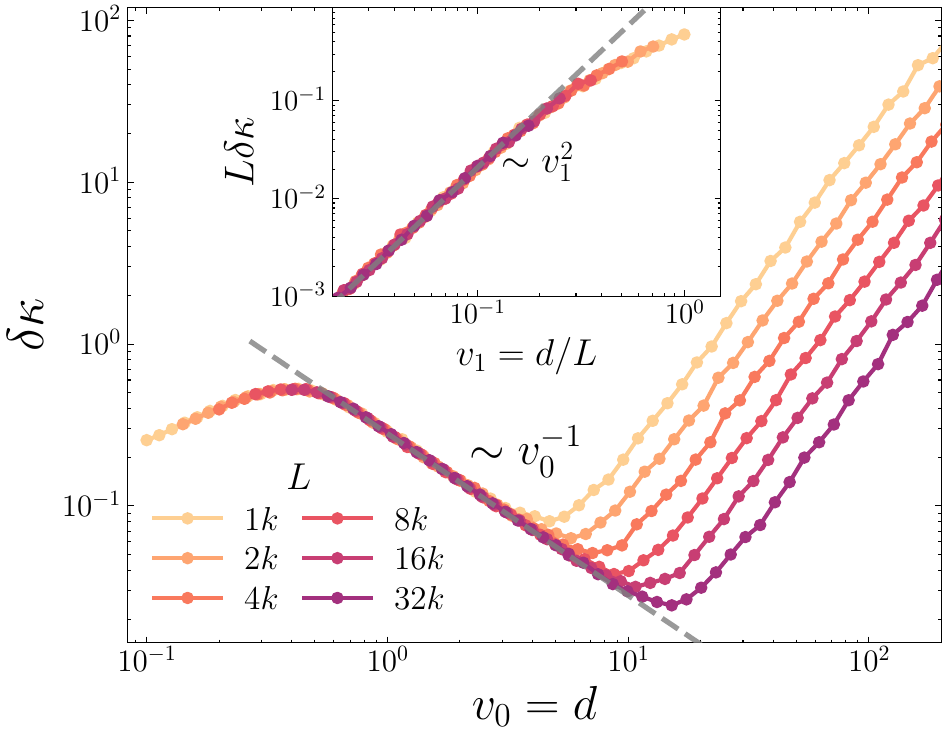}
\centering
\caption{{\bf Behavior of distribution width $\delta \kappa$ at extreme fluctuation scales.} Convergence towards the asymptotic scaling function $\Phi^{\delta \kappa}_{\gamma}(v_{\gamma})$, cf. Eq.~\eqref{eq:scaling_function_delta_kappa}, at extreme fluctuation scales $\gamma=0$ (main panel) and $\gamma=1$ (inset) for various system sizes, validating the scaling function $f_{\delta \kappa}=\Phi^{\delta \kappa}_{1/3}(v_{\gamma})$, cf. Eq.~\eqref{eq:scaling_function}, for $v_{0}\gg 1$ and $v_{1}\ll 1$. Dashed gray lines represent linear fitting functions.}
\label{fig:rescaling_dk}
\end{figure}

To establish the predicted algebraic exponents, namely $q(0\leq\gamma< 1/3)=-\gamma$ and $q(1/3\leq\gamma\leq1)=2\gamma-1$, we need to demonstrate that scaling functions $\Phi^{\delta \kappa}_{\gamma}$ obey
the asymptotics $\Phi^{\delta \kappa}_{\gamma}(v_\gamma)\sim v_\gamma^{-1}$ at small $v_{\gamma}$ for $0<\gamma<1/3$ and $\Phi^{\delta \kappa}_{\gamma}(v_\gamma)\sim v_\gamma^{2}$ at large $v_{\gamma}$ for $1/3<\gamma<1$, respectively. Our analysis confirms that this is the case in the entire interval $0<\gamma<1$ as exemplified for a representative value $\gamma=1/2$ in Fig.~\ref{fig:rescaling_dk}. The scaling function $\Phi^{\delta \kappa}_{1/3}(v_{1/3})$ connecting these two regimes is indeed the scaling function $f_{\delta \kappa}(v_{1/3})$ shown in Fig.~\ref{scaling-Vershik_b}.
Concerning the extreme fluctuation scales at $\gamma=0$ and $\gamma=1$, we find the scaling form
$\delta \kappa=L^{1/3}f_{\delta \kappa}(d/L^{1/3})$ to be still satisfied as long as $v_{0}\gg 1$ and $v_{1}\ll 1$, respectively, see Fig.~\eqref{fig:rescaling_dk} (while in the regimes $v_{0}\ll 1$ and $v_{1}\gg 1$ we observe systematic deviations). In conclusion, the scaling function $f_{\delta \kappa}$ accurately captures the scaling properties of $\delta \kappa$ in the entire regime $1\ll d\ll L$.

\paragraph*{Alternative sampling.}
We check whether the scaling exponents are robust with respect to the sampling procedure.
Previously, in Sec.~\ref{sec:Mod_Vershik}, we sampled states from a `$d$-vicinity' of a “perfect” discretization of the limit shape $\psi_{V}$. Here we try out an alternative sampling procedure:
we first draw a seed configuration $\lambda_{\mathrm{seed}}$ from the Vershik ensemble (representing a typical microstate from the Vershik macrostate), see Eq.~\eqref{eq:norm_sampling}, then generate a pair of eigenstates inside the $d$-vicinity of $\lambda_{\mathrm{seed}}$ according to
$\lambda_j' \sim \lfloor \lambda_{\mathrm{seed}} + d \cdot \, g_V \left( j / L \right)\rfloor,$
and finally compute the matrix element for such a pair. Afterwards, we draw a new seed configuration $\lambda_{\mathrm{seed}}$ and repeat the procedure to generate an ensemble of pairs of eigenstates.

\begin{figure}[H]
    \centering
    \includegraphics[width=1.0\linewidth]{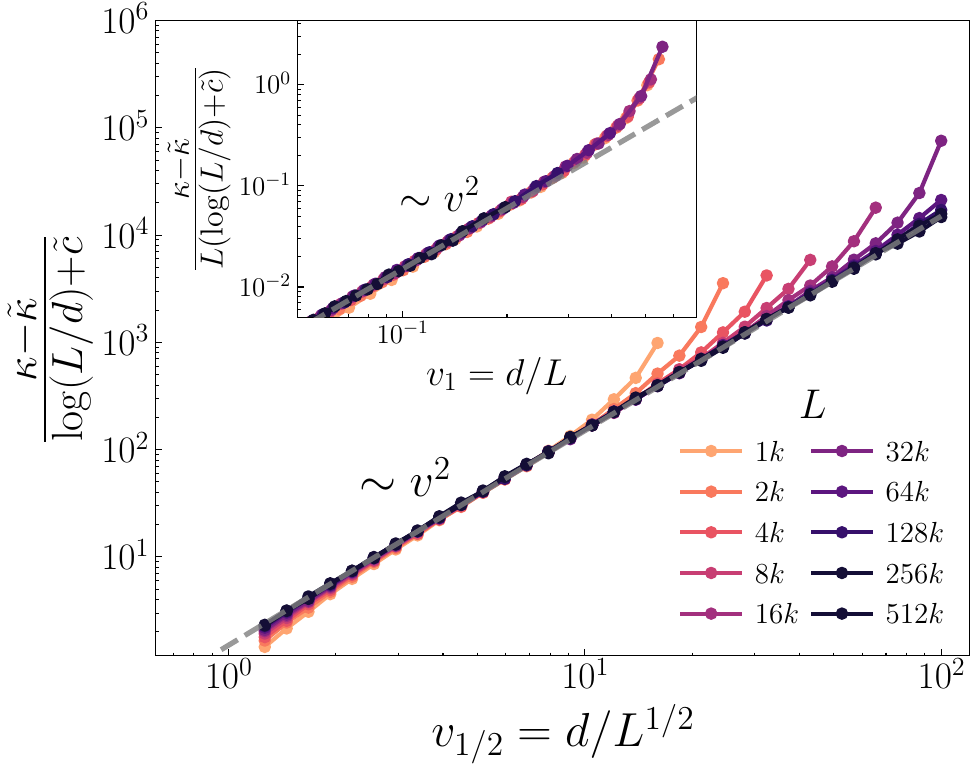}
    \caption{\textbf{Scaling analysis of suppression rate $\kappa$ using an alternative sampling.} Convergence towards the asymptotic scaling function $\Phi^{\kappa}_{\gamma}(v_{\gamma})$, cf. Eq.~\eqref{eq:scaling_function_kappa}, for fluctuation scales $\gamma=1/2$ (main panel) and $\gamma=1$ (inset) for various system sizes (including the subleading constant shifts $\tilde{\kappa}\approx 10$ and $\tilde{c}\approx -0.8$ to achieve better convergence) obtained with an alternative sampling protocol.
    Dashed gray lines represent linear fitting functions.}
    \label{fig:rescaling_k_non}
\end{figure}
\begin{figure}[H]
    \centering
    \includegraphics[width=1.0\linewidth]{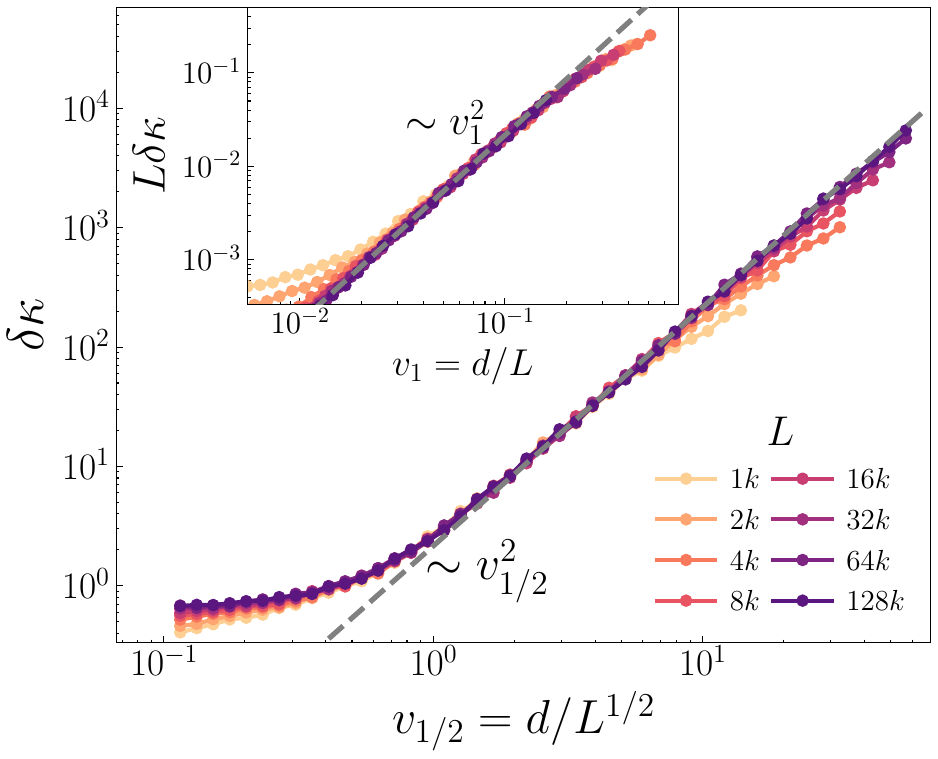}
    \caption{\textbf{Scaling analysis of distribution mean $\delta \kappa$ using an alternative sampling.}
    Convergence towards the asymptotic scaling function $\Phi^{\delta \kappa}_{\gamma}(v_{\gamma})$, cf. Eq.~\eqref{eq:scaling_function_delta_kappa}, for fluctuation scales $\gamma=1/2$ (main panel) and $\gamma=1$ (inset) for various system sizes obtained with an alternative sampling protocol.
    Dashed gray lines represent linear fitting functions.}
    \label{fig:rescaling_dk_non}
\end{figure}
Performing the same scaling analysis on the suppression rate $\kappa$ and width $\delta \kappa$ as previously with the “perfect” discretization, we again find the asymptotics $\Phi_{\gamma}^{\kappa}\sim v_{\gamma}^{2}$
holds in the interval $1/2<\gamma<1$, see Fig. \ref{fig:rescaling_k_non}. At $\gamma=1$ we likewise find
$\Phi_{1}^{\kappa}(v_1)$ to deviate from the $v^{2}_{\gamma}$ asymptotics for $v_1\gg 1$.
At the value of $\gamma=1/2$, $\Phi_{1/2}^{\kappa}\sim \tilde{\kappa}+v_{1/2}^{2}$, see Fig.~\ref{fig:rescaling_k_non}. In the interval $1/2\leq\gamma<1$ the suppression rate $\kappa$ therefore obeys $\kappa=O(L^{2\gamma-1}\log L)$, which in particular implies $p(\gamma)=2\gamma-1$. In the range $0<\gamma<1/2$ on the other hand, the $\tilde{\kappa}$ term dominates at large $L$, implying $\kappa=O(1)$ and thus $p(\gamma)=0$.

We finally analyze the scaling properties of the distribution width $\delta \kappa$. In the interval $1/2 < \gamma < 1$, our data is accurately described by the same scaling function, cf. Eq.~\eqref{eq:scaling_function_delta_kappa}. At the boundaries of this interval, i.e. at $\gamma=1/2$ and $\gamma=1$, we find the corresponding scaling functions to satisfy $\Phi^{\delta \kappa}_{1/2}(v_{1/2})\sim v_{1/2}^2$ for $v_{1/2}\gg 1$ and $\Phi^{\delta \kappa}_{1}(v_{1})\sim v_{1}^{2}$ for $v_{1}\ll 1$, respectively. Therefore, in the regime $1/2 \leq \gamma \leq 1$, the distribution width obeys $\delta \kappa = O(L^{2\gamma-1})$, so that $q(\gamma)=2\gamma-1$. By contrast, we have not found a scaling function that produces that collapses our data in the range $0<\gamma<1/2$. This might be due to the finite-size effects playing a prominent role, or perhaps a feature specific to the sampling ensemble.

\bibliographystyle{MyStyle}
\bibliography{references}

\end{document}